\documentclass{IEEEtran}
\usepackage{mathrsfs}
\usepackage[ruled,norelsize]{algorithm2e}
\usepackage{verbatim}
\ifCLASSINFOpdf
\else
\fi
\usepackage[cmex10]{amsmath}
\usepackage{amsthm,cite,graphicx,,booktabs,lipsum,color,bm,enumitem}
\usepackage{pifont,tikz,amssymb}
\usepackage[normalem]{ulem}
\theoremstyle{definition}

\makeatletter
\newcommand{\removelatexerror}{\let\@latex@error\@gobble}
\makeatother

\begin{document}
\title{Backhaul-aware Drone Base Station Placement and Resource Management for FSO-based Drone-assisted Mobile Networks}

\author{Liangkun~Yu,~\IEEEmembership{Graduate Student~Member,~IEEE,}
\and Xiang~Sun,~\IEEEmembership{Member,~IEEE,}
\and Sihua Shao,~\IEEEmembership{Member,~IEEE,}
\and Yougan Chen,~\IEEEmembership{Senior Member,~IEEE,}
 and Rana~Albelaihi,~\IEEEmembership{Graduate Student~Member,~IEEE}
\thanks{L. Yu, X. Sun, and R. Albelaihi are with the SECNet Lab., University of New Mexico, Albuquerque, NM 87131, USA. E-mail: $\{$liangkun,sunxiang, ralbelaihi$\}$@unm.edu. \par
S. Shao is with the Department of Electrical Engineering, New Mexico Tech, Socorro, NM 87801, USA. E-mail: sihua.shao@nmt.edu. \par
Y. Chen is with the Key Laboratory of Underwater Acoustic Communication and Marine Information Technology (Xiamen University), Ministry of Education, Xiamen 361005, China. E-mail: chenyougan@xmu.edu.cn \par
This work is supported by the National Science Foundation under Awards CNS-2148178 and OIA-1757207.\par
This work has been submitted to the IEEE for possible publication. Copyright may be transferred without notice, after which this version may no longer be accessible.}
}

\maketitle

\begin{abstract}

In drone-assisted mobile networks, Drone-mounted Base Stations (DBSs) are responsively and flexibly deployed over any Places of Interest (PoI), such as sporadic hotspots and disaster-struck areas, where the existing mobile network infrastructure is unable to provide wireless coverage. In this paper, a DBS is an aerial base station to relay traffic between a nearby Macro Base Station (MBS) and the users. In addition, Free Space Optics (FSO) is applied as the backhauling solution to significantly increase the capacity of the backhaul link between an MBS and a DBS. Most of the existing DBS placement solutions assume the FSO-based backhaul link provides sufficient link capacity, which may not be true, especially when a DBS is placed far away from an MBS (e.g., $>$ 10 km in disaster-struck areas) or in a bad weather condition. In this paper, we formulate a problem to jointly optimize bandwidth allocation and DBS placement by considering the FSO-based backhaul link capacity constraint. A \textbf{B}ackhaul awa\textbf{R}e bandwidth all\textbf{O}c\textbf{A}tion and \textbf{D}BS placement (BROAD) algorithm is designed to efficiently solve the problem, and the performance of the algorithm is demonstrated via extensive simulations. 
\end{abstract}

\IEEEpeerreviewmaketitle

\section{Introduction}
\begin{figure}[!htb]
	\centering	
 \includegraphics[width=1.0\columnwidth]{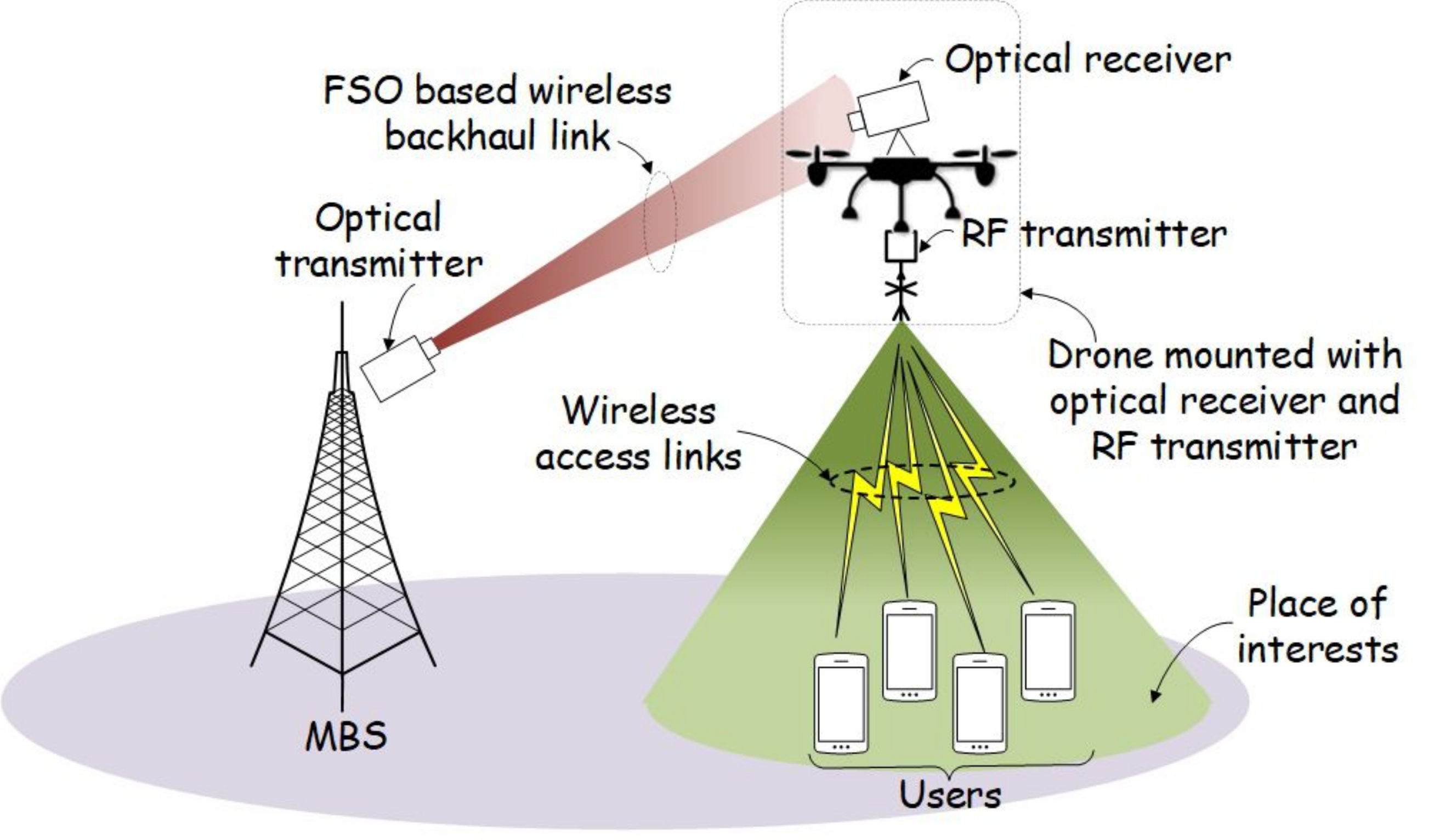} 
	\caption{FSO-based drone-assisted mobile network architecture.}
	\label{fig:main_archi}
\end{figure}

In drone-assisted mobile networks, Drone-mounted Base Stations (DBSs) are applied to be quickly and flexibly deployed over any Place of Interest (PoI), such as sporadic hotspots and disaster-struck areas, where the existing mobile network infrastructure is unable to provide wireless coverage or sufficient network capacity to the users \cite{Sun:MEC:2017,Zhang:2018:PMD, Chowdhery:2018:ACP}. Deploying a DBS over a PoI can assist a nearby Macro Base Station (MBS) in communicating with the users in the PoI. Here, a DBS is referred to as an aerial base station to relay traffic between an MBS and the users \cite{Sun:2017:LAD,wu2019cooperative,8938182}. Also, the capacity of the backhaul link between a DBS and an MBS may affect the throughput of the path between the MBS and the users via the DBS, and applying different wireless backhaul communications technologies lead to different backhaul link capacities and characteristics. For example, various Radio Frequency (RF) based backhaul solutions, such as microwave, mmWave, and sub-6 GHz, have been widely used in the current heterogeneous wireless networks \cite{Niu:2017:ESB,Chehri:2020:PMP,Coldrey:2013:NSC}.
However, these solutions may suffer from strong interference and severe attenuation \cite{Siddique:2015:WBS}, thus leading to low backhaul link capacity, especially when the distance between an MBS and a DBS is long (e.g., $>$~10~km). In this paper, we propose to use free space optics (FSO) as the backhaul solution to improve the backhaul link capacity in drone-assisted mobile networks. As shown in Fig. \ref{fig:main_archi}, an MBS sends an optical beam carrying the users' traffic to a DBS, which demodulates the received optical beam to retrieve the users' traffic and then forwards the traffic to the users in the PoI via the wireless access links \cite{9540913,9302228,9557287}.


As compared to RF, applying FSO as the backhaul solution has the following advantages. First, FSO can provide a higher link capacity over a longer distance between the two endpoints. It has been demonstrated that an FSO link can offer a $Gbps\!-\!Tbps$ data rate at the distance of several kilometers \cite{Ciaramella:2009:FSO,Curran:2017:FSONet,Esmail:2017:IDH}. Second, FSO is operated over unlicensed spectrum, which can reduce the operational cost of mobile providers \cite{Zhang:2016:FLU,Kaushal:2017:OCS,AMIRABADI2021165883}. Third, the spectrum used by FSO does not overlap with that used by the RF communications, which avoids the interference to/from terrestrial radio\cite{Trigui:2017:IMM}. Fourth, FSO is a type of secure communications since any interception and eavesdropping on an FSO link can be easily identified owing to the fact that an optical beam is directional with a narrow divergence angle \cite{Kaine:1995:IIT,Aveta:2018:ICA}.

The new architecture brings challenges to the DBS placement. 
First, FSO is a Line of Sight (LoS) communication, and the DBS placement should guarantee LoS between a DBS and an MBS. Second, the DBS placement is critical to determine the capacity of the network. Various DBS deployment methods have been designed, which only maximize the capacity of the access network between a DBS and the users in the PoI by assuming sufficient backhaul link capacity \cite{Zhang:2019:DBS,Wang:2019:ADU,Wu:2018:CTM}. 
This assumption may not be true, for example, in a disaster-struck scenario, when all the MBSs in a disaster-struck area could malfunction, and so deploying a DBS over the disaster-struck area to convey emergence communications may lead to a long distance (e.g., $>$ 10 km) between the DBS and a working MBS (which locates out of the disaster-struck area). The long distance between the MBS and the DBS may result in limited backhaul link capacity and constrain the network throughput. 
Also, the weather condition can have negative impacts on the FSO-based backhaul link capacity. 
For example, in thick foggy weather, the FSO-based backhaul link capacity will be tremendously reduced. Therefore, in order to maximize the overall throughput, it is critical to consider the capacities of both the backhaul link and access network in solving the DBS placement problem.

To resolve these challenges, this paper aims to design a joint backhaul-aware DBS placement and bandwidth allocation method in FSO-based drone-assisted mobile networks such that the number of satisfied users (i.e., the users whose data rate requirements are satisfied) is maximized, while guaranteeing the LoS between the MBS and the DBS. The contributions of the paper are summarized as follows.
\begin{enumerate}[leftmargin=*]
\item{We formulate the joint backhaul-aware DBS placement and bandwidth allocation problem in the context of FSO-based drone-assisted mobile networks.}
\item{We decompose the problem into two sub-problems, i.e., user access control and DBS placement updating, and design a heuristic algorithm, i.e., \textbf{B}ackhaul awa\textbf{R}e bandwidth all\textbf{O}c\textbf{A}tion and \textbf{D}BS placement (BROAD), to iteratively solve the two sub-problems.}
\item{The performance of BROAD is demonstrated via extensive simulations.}
\end{enumerate}

The rest of the paper is organized as follows. In Section II, we briefly introduce the related works. In Section III, the related system models 
are presented. In Section IV, we formulate the joint DBS placement and bandwidth allocation as an optimization problem. BOARD is designed to solve the problem in Section IV. Extensive simulations are explained and analyzed in Section V. A brief conclusion is drawn in Section VI.

\section{Related Works}
Researchers have shown an increased interest in resource management and DBS placement in drone-assisted mobile networks.  Al-Hourani \emph{et al.} \cite{Al-Hourani:2014:OLA} provided a probabilistic LoS pathloss model between ground users and a DBS, and designed a method to optimize the altitude of the DBS to maximize the size of the DBS's coverage area. Alzenad \emph{et al.} \cite{Alzenad:2017:PUA} designed a 3D DBS placement algorithm to maximize the number of covered users. The algorithm derives the optimal altitude of a DBS that maximizes the DBS's coverage area, and then adjusts the horizontal position of the DBS to cover the maximum number of users. Arribas \emph{et al.} \cite{Arribas:2019:FCT} designed a heuristic algorithm to optimize the bandwidth allocation, user association, and multi-DBS placement to maximize the $\alpha$-fair throughput utility function, which is used to measure the tradeoff between fairness and throughput in the access network. Other works investigated the DBS placement in the context of FSO-based drone-assisted mobile networks, where FSO is considered as the backhaul solution. However, \emph{they all assumed that the capacity of the FSO-based backhaul link is sufficient to accommodate the traffic demand of the access network, which is impractical in some scenarios}. For example, Sun \emph{et al.} \cite{Sun:2019:JOD} designed an algorithm to determine the 3D position of a DBS and the user association to maximize the overall Spectrum Efficiency (SE) of a PoI in the access network. Specifically, in each iteration, the designed algorithm sequentially derives the user association, horizontal position, and altitude of the DBS that can increase the SE of the PoI. The iteration terminates once the SE of the PoI cannot be further improved. Zhang and Ansari \cite{Zhang:2019:DBS} jointly optimized the 3D DBS placement, bandwidth allocation, and power management to maximize the overall throughput of the access links, while guaranteeing the data rate requirement of the users. They also explored the relationship among users' QoS requirement, user association, and bandwidth allocation \cite{9537707}. Di \emph{et al.} \cite{Wu:2019:FDA} designed a heuristic algorithm to optimize the multi-DBS deployment, user association, and bandwidth allocation in disaster-struck scenarios, where all the MBSs in a disaster-struck area are damaged and DBSs are deployed over the area to provide emergency communications.

Some works explored the backhaul-aware DBS placement in drone-assisted mobile networks, where traditional RF communications are applied as the backhaul solution. Kalantari \emph{et al.} \cite{Kalantari:2017:BRD} designed a heuristic algorithm to determine DBS placement and bandwidth allocation such that the number of users, whose pathloss to the DBS is larger than the predefined threshold, can be maximized, while guaranteeing the overall throughput of access network no larger than the capacity of the wireless backhaul link. However, they assumed that the capacity of the backhaul link does not change by varying the DBS placement, which may not be a practical assumption. Sun and Ansari \cite{Sun:2018:JOD} assumed that the DBS is operated in the in-band half-duplex mode, i.e., backhaul and access link communications are conducted over the same frequency band but in different time slots. They designed a heuristic algorithm to optimize the DBS placement and user association to maximize the overall spectral efficiency in a PoI. Zhang \emph{et al.} \cite{Zhang:2018:DPI} proposed a heuristic algorithm to maximize the overall throughput by optimizing the DBS placement, and bandwidth and power allocation in both backhaul and access links, where the DBS is operated in the in-band full-duplex mode.

\section{System Models}
 \begin{table}[!htb]

	\caption{Notations}
	\begin{tabular}{ll}
		\toprule
		\textbf{Notation} & \textbf{Definition} \\
		\midrule
		$\bm{\mathcal{I}}$ & Set of ground users in the PoI\\
            $d_i$ &3D distance between the DBS and user $i$\\
            $\eta_i$ & Pathloss between the DBS and user $i$\\
            $f_c$ &  Carrier frequency in the wireless access network\\
            $\rho_i$ & Probability of having LoS between the DBS and user $i$\\
            $\xi ^{los}$/$\xi ^{nlos}$ &Average additional pathloss for LoS/NLoS\\
            $\theta_i$ & Elevation angle between the DBS and user $i$\\
            $p^{fso}$ & Transmission power of the FSO transmitter at the MBS\\
	        $\tau^{tx}$/$\tau^{rx}$&	Optical efficiency of the FSO transmitter/receiver \\
            $\gamma$ & Atmospheric attenuation factor\\
            $v$ & Visibility distance  \\
            $\lambda$ & Wavelength of the optical beam \\
		    $L $ & 3D distance between the MBS and the DBS\\
            $\vartheta$ & Diameter of the optical receiver’s aperture\\
	        $\varepsilon$ & Divergence angle of FSO transmitter\\
	        $E_p$ & Photon energy at wavelength $\lambda$\\
	        $\kappa$ & Planck constant \\
            $c$ & Speed of light\\
	        $N_b$ & Sensitivity of the FSO receiver at the DBS  \\
		\bottomrule
	\end{tabular}%
  \label{tab_models}
\end{table}%
The major notations in the system models and problem formulation are listed Table \ref{tab_models}. Assume that a DBS will be placed over or near a PoI to assist the users in the PoI in downloading traffic from an MBS via the DBS. Denote $(x, y)$ as the 2D coordinates of the DBS on a horizontal plane. Denote $h$ as the altitude of the DBS. 
Let $\bm{\mathcal{I}}$ be the set of users in the PoI and $i$ be the index of these users.
Denote $(x_{i},y_{i})$ as the 2D coordinates of user $i$. Also, denote $\varphi_i$ as user $i$'s data rate requirement. Thus, the 3D distance between the DBS and  user $i$ is
\begin{equation}
    {d_i} = \sqrt {l_i^2 + {h^2}},
    \label{eq:3d_dist}
\end{equation}
where $l_{i}=\sqrt{(x-x_i)^2+(y-y_i)^2}$ is the horizontal distance between user $i$ and the DBS.

\subsection{Pathloss model between the DBS and the users}
The wireless propagation channel between a DBS and a user can be divided into two scenarios, i.e., the link between a DBS and a user with LoS and Non-Line of Sight (NLoS) connections \cite{Al-Hourani:2014:OLA, Zhang:2018:MLP, Ghanavi:2018:EAB}. The pathloss in LoS is lower than that in NLoS since the signals from the DBS may suffer from much stronger reflections and diffraction in NLoS \cite{Khawaja:2019:SAG,8688470}. Thus, the average pathloss (in dB) between a DBS and user $i$ can be estimated by
\begin{equation}
{\eta_i} = 20{\log_{10}}\left( {\frac{{4\pi {f_c}{d_i}}}{c}} \right) + {\rho _i}{\xi ^{los}} + \left( {1 - {\rho _i}} \right){\xi ^{nlos}},
\label{eq:DBS_pathloss}
\end{equation}
where $f_c$ is the carrier frequency, $d_i$ is the 3D distance between the DBS and user $i$, $c$ is the speed of light, $\rho_i$ is the probability of having LoS between the DBS and user $i$, and $\xi ^{los}$ and $\xi ^{nlos}$ are the average additional pathloss for the LoS and NLoS scenarios, respectively. Here, $20\log_{10}\left( {\frac{{4\pi {f_c}{d_i}}}{c}} \right)$ indicates the free space pathloss between the DBS and user $i$, and ${\rho _i}{\xi ^{los}} + \left( {1 - {\rho _i}} \right){\xi ^{los}}$ is the average additional pathloss between the DBS and user $i$. The probability of having LoS between the DBS and user $i$ (i.e., $\rho_i$) can be estimated by \cite{al2014optimal}
\begin{equation}
{\rho_{i}}\!=\!\frac{1}{{1\!+\!\alpha {e^{ -\!\beta \left( {{\theta _i}\!-\!\alpha } \right)}}}}\!=\!\frac{1}{{1\!+\!{\alpha}{e^{\!-\!{\beta}\!\left(\!{\frac{{180}}{\pi }{\arctan \left( {\frac{h}{l_i}} \right)}\!-\!{\alpha}}\!\right)}}}},
\label{eq:los_prob}
\end{equation}
where $\theta _i$ (in degrees) is the elevation angle between the DBS and user $i$, and $\alpha$ and $\beta$ are the environmental parameters determined by the environment of the hotspot area (e.g., rural, urban, etc.). 
\subsection{Access link data rate model}
The achievable data rate of downloading data streams from the DBS to users $i$ is 
\begin{equation}
r_i^{access} = {b_i}{\log _2}\left( {1 + \frac{{p^d}{{10}^{-{\textstyle{{{\eta _i}} \over {10}}}}}}{N_0}} \right), 
\label{eq:data_rate_model}
\end{equation}
where $b_i$ is the amount of bandwidth assigned to user $i$, $p^d$ is the transmission power of the DBS, $N_{0}$ is noise power level, and ${10}^{-{\textstyle{{{\eta _i}} \over {10}}}}$ is the channel gain between the DBS and user $i$\footnote{For a clear exposition, the shadowing and fading effects are not considered in calculating the channel gain.}. Here, $\eta _i$ is the average pathloss between the DBS and user $i$, which can be estimated based on Eq. \eqref{eq:DBS_pathloss}. From Eq. \eqref{eq:data_rate_model}, it is easy to derive that the achievable data rate of user $i$ depends on the amount of allocated bandwidth and the pathloss, which is determined by the DBS placement.

\subsection{FSO-based backhaul link data rate model}
\begin{figure}[!htb]
	\centering	
 \includegraphics[width=1.0\columnwidth]{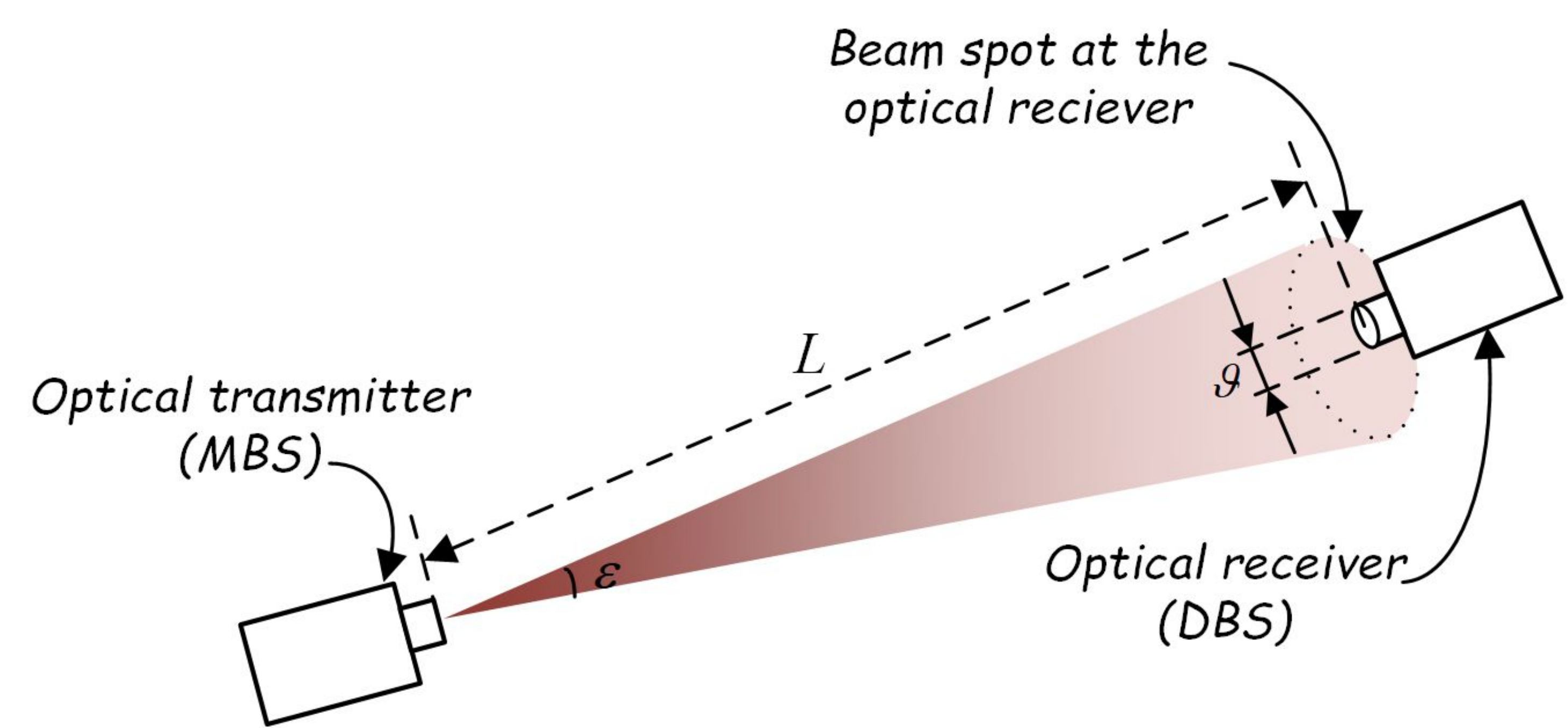} 
	\caption{Illustration of the FSO-based backhaul link.}
	\label{fig:optical_beam}
\end{figure}

FSO is applied to achieve the backhaul communications between an MBS and a DBS. Note that the DBS placement must be able to establish LoS communication with the MBS, which is a critical pre-requirement for FSO. In this paper, we assume that if the DBS is placed above a predefined altitude, denoted as $h^{min}$, LoS can be maintained between the MBS and the DBS. Under the LoS condition, the achievable data rate of an FSO link can be estimated by \cite{Hu:2007:FOC}
\begin{equation}
{r^{fso}} = \frac{{p^{fso}{\tau^{tx}}{\tau^{rx}}{{10}^{ - \frac{{\gamma L}}{{10}}}}{\vartheta ^2}}}{{\pi {{\left( {{\varepsilon}/2} \right)}^2}{L^2}{E_p}{N_b}}},
\label{eq:fso_data_rate}
\end{equation}
where $p^{fso}$ is the transmission power of the FSO transmitter at the MBS, $\tau^{tx}$ is the optical efficiency of the FSO transmitter, $\tau^{rx}$ is the optical efficiency of the FSO receiver at the DBS, $\vartheta$ is the diameter of the FSO receiver's aperture shown in Fig. \ref{fig:optical_beam}, $\varepsilon$ is the divergence angle of the FSO transmitter, $E_p$ is the photon energy at wavelength $\lambda$ (i.e., $E_p=\kappa c/\lambda$, where $\kappa$ is Planck constant, $c$ is the speed of light, and $\lambda$ is the photon's wavelength), $N_b$ is the FSO receiver sensitivity, and $L$ is the 3D distance between the MBS and the DBS, i.e.,
\begin{equation}
L = \sqrt {{{\left( {x - {x^m}} \right)}^2} + {{\left( {y - {y^m}} \right)}^2} + {{\left( {h - {h^m}} \right)}^2}},
\end{equation}
where $\left( {{x^m},{y^m},{h^m}} \right)$ is the 3D coordinates of the MBS. $\gamma$ in Eq. \eqref{eq:fso_data_rate} is the atmospheric attenuation factor in $dB/km$, which is determined by the visibility distance $v$ (i.e., the maximum distance that one object can be clearly discerned) and the size distribution of the scattering particles $q$ \cite{6843013}, i.e., 
\begin{equation}
    \gamma=\frac{3.91}{v}(\frac{\lambda}{550})^{-q}.
\label{eq:visibiltity_range}
\end{equation} 
Note that the visibility distance $v$ depends on the current weather condition. For example, the visibility distance on a clear day could be more than $v=20$ km but may be less than $v\le1$ km on a foggy day. The value of $q$ can be estimated according to $v$ based on the following equation \cite{6843013},
\begin{equation}
\label{visibility distance}
    q=\left\{\begin{matrix}
1.6, & 50<v.\\ 
1.3, & 6<v\leq50.\\ 
 0.16v+0.34,& 1<v\leq6.\\ 
v-0.5, & 0.5<v\leq 1.\\ 
0, & v\leq0.5.
\end{matrix}\right.
\end{equation}
 
In general, if the DBS is placed closer to the MBS (i.e., a smaller $L$), the FSO-based backhaul link can achieve a higher data rate, and vice versa. Note that the achievable data rate of an FSO link estimated by Eq. \eqref{eq:fso_data_rate} does not consider the pointing loss owing to the fact that various technologies, such as high-accuracy Acquisition, Tracking, and Pointing (ATP) systems \cite{Sun:2022:LCA,Liu:2020:DED} and adaptive probabilistic constellation shaping \cite{Fernandes:2022:FST}, have been proposed to significantly mitigate the pointing loss. However, a more sophisticated model \cite{4267802}, which takes both pointing loss and atmospheric turbulence into consideration, can be applied to estimate the achievable data rate of an FSO link. Yet, changing the achievable data rate model does not affect the proposed problem formulation in Section IV and the designed algorithm in Section V.  

\section{Problem Formulation}
We formulate the joint DBS placement and bandwidth allocation (in the access network) problem to maximize the number of the \emph{satisfied users} as follows.
Here, a \emph{satisfied user} is referred to as a user whose achievable data rate is no less than its data rate requirement.
\begin{align}
\bm{P0}:&\mathop {\arg \max }\limits_{{x,y},h,\bm{z}} \sum\limits_{i \in \bm{\mathcal{I}}} {z_i},\\
\text{s.t.}\ \ \ \ 
& B\ge \sum\limits_{i \in \bm{\mathcal{I}}} {{b_i}z_i}  , \label{const:band}\\
&{r^{fso}} \ge \sum\limits_{i \in \bm{\mathcal{I}}} r_i^{access}z_i  , \label{const:backhaul} \\
&\forall i \in \bm{\mathcal{I}},r_i^{access}{z_i} \ge {\varphi _i}{z_i}, \label{const:rate_req} \\
&{h^{\min }}  \le h \le h^{\max}, \label{const:altitude}\\
&\forall i \in \bm{\mathcal{I}}, {z_i} \in \left\{ {0,1} \right\},\label{const:integer_var}
\end{align} 
where $\bm{z} = \left\{ {{z_i}\left| {\forall i \in \bm{\mathcal{I}}} \right.} \right\}$, $z_i$ is a binary variable to indicate whether user $i$ is a \emph{satisfied user} (i.e., $z_i=1$) or not (i.e., $z_i=0$), $B$ is the total amount of bandwidth available for the access network, $b_i$ is the bandwidth assigned to user $i$, $\varphi _i$ is the data rate requirements of user $i$, $h^{min}$ is the minimum altitude for the DBS to guarantee the LoS connection to the MBS\footnote{Note that different horizontal locations may have different minimum height requirements to ensure LoS between the MBS and the DBS. For instance, if location $(x_1, y_1)$ is much closer to the MBS than location $(x_2, y_2)$, then $h_{(x_1, y_1)}^{min}$ for location $(x_1, y_1)$ might be lower than $h_{(x_2, y_2)}^{min}$ for location $(x_2, y_2)$. Here, the minimum height of any location $(x, y)$, denoted as $h_{(x, y)}^{min}$, can be calculated based on the height of the MBS and the heights of the buildings/obstacles between the MBS and location $(x, y)$. The minimum height $h^{min}$ of the DBS in Constraint (13) equals the maximum value of all the minimum heights for all the possible locations for the DBS, i.e., $h^{\min}=\max \left\{ h_{\left( x,y \right)}^{\min}\left| \left( x,y \right) \in \bm{R} \right. \right\}$, where $\bm{R}$ represents the PoI area plus the horizontal area between the MBS and the DBS.}, and $h^{max}$ is the maximum altitude that the DBS can reach. Constraint \eqref{const:band} indicates that the amount of bandwidth allocated to the users in the PoI should be no larger than the total amount of available bandwidth. Constraint \eqref{const:backhaul} means that the achievable data rate of the FSO-based backhaul link should be no less than the sum of the achievable data rates from the DBS to the satisfied users in the PoI. Constraint \eqref{const:rate_req} implies that, for each \emph{satisfied user}, its achievable data rate should be no less than its data rate requirement. Constraint \eqref{const:altitude} defines the minimum and maximum altitude of the DBS. 

\section{Backhaul-aware bandwidth allocation and DBS placement}
In order to efficiently solve $\bm{P0}$, we design the \textbf{B}ackhaul awa\textbf{R}e bandwidth all\textbf{O}c\textbf{A}tion and \textbf{D}BS placement (BROAD) algorithm. The basic idea of BROAD is to decompose $\bm{P0}$ into two sub-problems, i.e., user access control (denoted as $\bm{P1}$) and DBS placement (denoted as $\bm{P2}$), and iteratively solve the two sub-problems until the number of the \emph{satisfied users} cannot be further increased.

Note that, to maximize the number of satisfied users, we need to minimize the allocated bandwidth to the satisfied users, while satisfying their data rate requirements. So, it is straightforward to allocate the exact amount of bandwidth to a satisfied user such that its achievable date rate just equals its data rate requirement, i.e., $r^{access}_i=\varphi_i$. That is,
\begin{equation}
    b_i=\frac{{{\varphi _i}}}{{{{\log }_2}\left( {1 + \frac{{{p^d}{{10}^{-{\textstyle{{{\eta _i}} \over {10}}}}}}}{{{N_0}}}} \right)}}.
\end{equation} 

\subsection{User access control}
Assume that the 3D coordinates of the DBS (i.e., $\left( {x,y,h} \right)$) are given. Then, $\bm{P0}$ can be converted into $\bm{P1}$, where
\begin{align}
\bm{P1}:
&\mathop {\arg \max }\limits_{\bm{z}}\sum\limits_{i \in \bm{\mathcal{I}}} z_i\\
\text{s.t.}\ \ \ \ 
&r^{fso} \ge \sum\limits_{i \in {\bm{\mathcal{I}}}} {{\varphi _i}z_i}   , \label{const:backhaul_fi} \\
&\text{Constraints}\ \eqref{const:band},\eqref{const:integer_var}.
\end{align}
Basically, $\bm{P1}$ is to determine which users will be selected and allocated with enough bandwidth to meet their data rate requirements such that the number of satisfied users is maximized. Note that $\bm{P1}$ can be mapped into a special 0-1 multi-dimension Knapsack problem, where we have to determine which items have to be collected into a knapsack such that the total value of the items in the knapsack is maximized, and the total size and weight of the items in the knapsack should be no larger than the size and weight capacity of the knapsack, respectively. Here, $\varphi_i$, $b_i$, $f^{fso}$, and $B$ in $\bm{P1}$ are considered as the size of item $i$, weight of item $i$, size capacity of the knapsack, and weight capacity of the knapsack, respectively. $z_i$ is a binary variable to indicate whether item $i$ should be put into a knapsack or not. Note that the values of all the items are the same equal to 1. The genetic algorithm (GA) \cite{shah2019genetic,chu1998genetic} is applied to solve $\bm{P1}$. GA comprises the following steps.
\begin{enumerate}[leftmargin=*]
\item{Initially, GA randomly generates $n$ feasible solutions for $\bm{P1}$. Denote $\bm{\mathcal{K}}$ as the set of the feasible solutions, and $k$ is used to index these feasible solutions. Also, let ${\bm{\mathcal{Z}}^k} = \left\{ {z_1^k,z_2^k, \cdots z_{\left| \bm{\mathcal{I}} \right|}^k} \right\}$ be a feasible solution $k$ in $\bm{\mathcal{K}}$. Denote $f$ as the maximum objective value of $\bm{P1}$ among these feasible solutions, i.e.,
\begin{equation}
f = \mathop {\max }\limits_{1 \le k \le \left| \bm{\mathcal{K}} \right|} \left\{ {\sum\limits_{i \in \bm{\mathcal{I}}} {z_i^k} } \right\}.
\label{eq:GA_stop_condi}    
\end{equation}}
\item{All the feasible solutions in $\bm{\mathcal{K}}$ are equally separated into two sets, denoted as $\bm{\mathcal{K}}_1$ and $\bm{\mathcal{K}}_2$. Denote $k_1$ and $k_2$ as the indices of the feasible solutions that incur the largest objective value of $\bm{P1}$ in $\bm{\mathcal{K}}_1$ and $\bm{\mathcal{K}}_2$, respectively, i.e., ${k_1} = \mathop {\arg \max }\limits_{k \in \bm{\mathcal{K}}_1} \sum\limits_{i \in {\bm{\mathcal{I}}}} {z_i^k}$ and ${k_2} = \mathop {\arg \max }\limits_{k \in \bm{\mathcal{K}}_2} \sum\limits_{i \in {\bm{\mathcal{I}}}} {z_i^k}$.}
\item{The selected feasible solutions ${\bm{\mathcal{Z}}_{k_1}}$ and ${\bm{\mathcal{Z}}_{k_2}}$ are used to explore $m$ new solutions based on the crossover and mutation processes. Specifically, in the crossover process, each element in a new solution, denoted as ${\bm{\mathcal{Z}}^{k'}} = \left\{ {z_1^{k'},z_2^{k'}, \cdots z_{\left| \bm{\mathcal{I}} \right|}^{k'}} \right\}$, is generated by randomly selecting the corresponding element either in ${\bm{\mathcal{Z}}_{k_1}}$ or ${\bm{\mathcal{Z}}_{k_2}}$. That is, $\forall i \in \bm{\mathcal{I}}$, $z_i^{k'}=rand\left\{ {z_i^{{k_1}},z_i^{{k_2}}} \right\}$, where $rand\left\{  \bullet  \right\}$ is a function which randomly picks one value among the values defined inside the bracket. Based on the crossover process, $m$ new solutions will be created. In the mutation process, $q$ elements in each new solution will be randomly selected and flipped their values. For example, assume that $q=2$, and then two elements in new solution ${\bm{\mathcal{Z}}^{k'}}$ are randomly selected. Assuming that the 3$^{rd}$ and 10$^{th}$ elements are selected and their values are flipped, i.e., $z_3^{{k'}}=\left| {1 - z_3^{k'}} \right|$ and $z_{10}^{{k'}}=\left| {1 - z_{10}^{k'}} \right|$. Denote $\bm{\mathcal{K}}^{new}$ as the generated new solution set after the crossover and mutation processes.}
\item{The generated $m$ new solutions in $\bm{\mathcal{K}}^{new}$ may not be the feasible solutions to satisfy Constraints \eqref{const:backhaul_fi} and \eqref{const:band} in $\bm{P1}$. Denote $\bm{\mathcal{K}}^{inf}$ as the set of infeasible solutions in $\bm{\mathcal{K}}^{new}$. These infeasible solutions should be repaired to become feasible. Specifically, for an infeasible solution, denoted as ${\bm{\mathcal{Z}}^{k'}} = \left\{ {z_1^{k'},z_2^{k'}, \cdots z_{\left| \bm{\mathcal{I}} \right|}^{k'}} \right\}$ (where $k'$ is assumed to be the index of the infeasible solution), each element $i$ (i.e., user $i$) is associated with a utility ratio $\zeta_i$, i.e.,
\begin{equation}\label{eq_ga2}
    \zeta_i=\frac{2}{l_1\varphi_i+l_2b_i},
\end{equation}
where $l_1$ and $l_2$ are the two solutions of the dual problem corresponding to the relaxed $\bm{P1}$ (where binary variable $z_i$ is relaxed into a continuous variable) \cite{pirkul1987heuristic}, and $\varphi_i$ and $b_i$ are the data rate requirement and amount of bandwidth allocation to element $i$ (i.e., user $i$), respectively. Then, in each round, the element, which incurs the largest utility ratio $\zeta_i$ among all the elements whose values are equal 1 in ${\bm{\mathcal{Z}}^{k'}}$, will flip its value to 0. That is, $z_{{i^*}}^{k'} = 0$, where ${i^*} = \mathop {\arg \max }\limits_{i \in {\bm{\mathcal{I}}}} \left\{ {{\zeta _i}\left| {z_i^{k'} = 1} \right.} \right\}$. If solution ${\bm{\mathcal{Z}}^{k'}}$ after the element flipping still cannot meet Constraints \eqref{const:backhaul_fi} and \eqref{const:band}, GA goes to the next round to flip the value of the element in ${\bm{\mathcal{Z}}^{k'}}$. The round continues to flip the value of the element until ${\bm{\mathcal{Z}}^{k'}}$ can meet Constraints \eqref{const:backhaul_fi} and \eqref{const:band}. The whole repairing process ends once all the infeasible solutions become feasible, i.e., $\bm{\mathcal{K}}^{inf}=\emptyset$.}
\item{The generated new feasible solutions will be added into $\bm{\mathcal{K}}$, i.e., $\bm{\mathcal{K}}=\bm{\mathcal{K}} \cap \bm{\mathcal{K}}^{new}$. The maximum objective value $f$ is updated based on $\bm{\mathcal{K}}$. If $f$ in the current iteration does not increase as compared to that in the previous iteration, then GA terminates; otherwise, GA starts the next iteration to generate new feasible solutions based on Steps 2-4.}
\end{enumerate}

\subsection{DBS placement}
The termination of the user access control algorithm indicates that the utilization of the backhaul link and/or access network reach to 100\%, (i.e., Constraints \eqref{const:backhaul_fi} and/or \eqref{const:band} will be violated by adding any new satisfied user). Hence, DBS placement is applied to adjust the position of the DBS such that the utilization of the path from the MBS to the existing satisfied users via the DBS is minimized. Here, the utilization of the path equals the maximum value between the utilization of the FSO-based backhaul link (i.e., $\sum\limits_{i \in  {\bm{\mathcal{I}}}}\frac{z_i\varphi _i}{r^{fso}}$) and the access network (i.e., $\sum\limits_{i \in  {\bm{\mathcal{I}}}}\frac {z_ib_i}{B}$). That is, $\textit{ max}\left\{\sum\limits_{i \in  {\bm{\mathcal{I}}}}\frac{z_i\varphi _i}{r^{fso}},\sum\limits_{i \in  {\bm{\mathcal{I}}}}\frac {z_ib_i}{B}\right\} $. Reducing the utilization of the path implies that more satisfied users can be added by user access control in the next iteration. We then formulate the DBS placement problem as follows.
\begin{align}
\bm{P2}:& \mathop {\arg \min }\limits_{x,y,h}\textit{ max}\left\{\sum\limits_{i \in  {\bm{\mathcal{I}}}}\frac{z_i\varphi _i}{r^{fso}},\sum\limits_{i \in  {\bm{\mathcal{I}}}}\frac {z_ib_i}{B}\right\} \label{P2_obj_1},\\
\text{s.t.}\ \ 
&\text{Constraints}\ \eqref{const:altitude}, \eqref{const:band}, \text{and}\ \eqref{const:backhaul_fi}.\nonumber
\end{align}
Since the objective function Eq. \eqref{P2_obj_1} is not continuous and differentiable (which makes the problem difficult to be solved), we introduce an auxiliary variable $\varsigma$, where
\begin{equation}
\varsigma= \textit{ max}\left\{\sum\limits_{i \in  {\bm{\mathcal{I}}}}\frac{z_i\varphi _i}{r^{fso}},\sum\limits_{i \in  {\bm{\mathcal{I}}}}\frac {z_ib_i}{B}\right\}.
\end{equation}
Therefore, Constraints \eqref{const:band} and \eqref{const:backhaul_fi} in $\bm{P2}$ can be transformed into $\varsigma \le 1$. Accordingly, $\bm{P2}$ can be transformed into,
\begin{align}
\bm{P2\_1}:& \mathop {\arg \min }\limits_{x,y,h,\varsigma}\varsigma \label{P2_obj}\\
\text{s.t.}\ \ 
&\varsigma \ge \sum\limits_{i \in  {\bm{\mathcal{I}}}}\frac{z_i\varphi _i}{r^{fso}},\label{p21_fso}\\
&\varsigma \ge \sum\limits_{i \in  {\bm{\mathcal{I}}}}\frac {z_ib_i}{B},\label{p21_bandwidth}\\
 &\varsigma \le 1 ,\\
&h\le h^{\max},\\
&h\ge h^{\min}.
\end{align}
$\bm{P2\_1}$ is not a convex problem as Constraints \eqref{p21_fso} and \eqref{p21_bandwidth} are not convex. Sequential Quadratic Programming (SQP)\cite{nocedal2006numerical} is proposed to efficiently derive the local optimal of $\bm{P2\_1}$. SQP is a method to solve non-convex optimization problems, where the objective function and the constraints are continuous and twice differentiable. The idea of SQP is to iteratively construct and solve a quadratic programming sub-problem until the algorithm finds a local optimal. Here, the quadratic programming sub-problem is an approximation to $\bm{P2\_1}$ at point ${{\bf{u}}^{\left( t \right)}}=\left( {{x^{(t)}},{y^{(t)}},{h^{(t)}},{\varsigma^{(t)}}} \right)$, where $t$ indicates the $t^{th}$ iteration. Specifically, denote $\mathcal{L}\left({\bf{u}},\bf{m}\right)$ as the Lagrangian function of $\bm{P2\_1}$, where 
\begin{align}
&\mathcal{L}\left({\bf{u}},\bf{m}\right)\!=\varsigma\!-\!{m_1}\left( \varsigma - \sum\limits_{i \in  {\bm{\mathcal{I}}}}\frac{z_i\varphi _i}{r^{fso}} \right)\!-\! {m_2}\!\left( \varsigma - \sum\limits_{i \in  {\bm{\mathcal{I}}}}\frac {z_ib_i}{B}\right) \nonumber\\
&\!-\! {m_3}\!\left( 1-\varsigma \right)\!-\! {m_4}\!\left( {{h^{\max }}\!-\!{h}} \right)\!-\! {m_5}\!\left( {{h}\!-\!{h^{\min }}} \right).
\end{align}
Here, ${{\bf{m}}} = \left( {m_1,m_2,m_3,m_4,m_5} \right)$ are the Lagrangian multipliers corresponding to the constraints in $\bm{P2\_1}$. 
Also, let ${\bf{H}}^{\left( t \right)}$ be the Hessian matrix of the Lagrangian function $\mathcal{L}\left({\bf{u}},\bf{m}\right)$ for $\bm{P2\_1}$ at point ${{\bf{u}}^{\left( t \right)}}$. Then, we can construct the following quadratic programming sub-problem, which basically reflects the local properties of $\bm{P2\_1}$, at point ${{\bf{u}}^{\left( t \right)}}$.   

\begin{align}
& \bm{P3}:
\mathop {\arg \min }\limits_{{\bf{\Delta }}{{\bf{u}}^{\left( t \right)}}} \varsigma^{(t)}+ \Delta \varsigma^{(t)}+\frac{1}{2}{\bf{\Delta }}{{\bf{u}}^{\left( t \right)}}^T\!\!{\bf{H}}^{\left( t \right)}{\bf{\Delta }}{{\bf{u}}^{\left( t \right)}}, \label{P3_const1}\\
\text{s.t.}
&\left(\!\!\varsigma^{(t)}\!-\!\sum\limits_{i \in  {\bm{\mathcal{I}}}}\frac{z_i\varphi _i}{r^{fso}\left(\!{\bf{u}}^{\left( t \right)}\!\right)}\!\!\right)\!+\!\nabla\!\!\left(\!\!\varsigma^{(t)}\!-\!\sum\limits_{i \in  {\bm{\mathcal{I}}}}\frac{z_i\varphi _i}{r^{fso}\left(\!{\bf{u}}^{\left( t \right)}\!\right)}\!\!\right)^T\!\!\!\!\!{\bf{\Delta }}{{\bf{u}}^{\left( t \right)}}\!\ge\!0,\label{p3_1}\\
&\left(\!\!\varsigma^{(t)}\!-\!\sum\limits_{i \in  {\bm{\mathcal{I}}}}\frac{z_ib_i\left(\!{\bf{u}}^{\left( t \right)}\!\right)}{B}\!\!\right)\!+\!\nabla\!\!\left(\!\!\varsigma^{(t)}\!-\!\sum\limits_{i \in  {\bm{\mathcal{I}}}}\frac{z_ib_i\left(\!{\bf{u}}^{\left( t \right)}\!\!\right)}{B}\right)^T\!\!\!\!\!{\bf{\Delta }}{{\bf{u}}^{\left( t \right)}}\!\ge\!0,\label{p3_2}\\ 
&1-(\varsigma^{(t)}+\Delta \varsigma^{(t)})\ge 0,\label{p3_3}\\
&h^{max}-\left(h^{(t)}+\Delta h^{(t)} \right)\ge0,\label{p3_4}\\
&\left(h^{(t)}+\Delta h^{(t)}\right) -h^{min}\ge0,\label{p3_5}
\end{align}
where 
\begin{align}
    {\bf{\Delta }}{{\bf{u}}^{\left( t \right)}}&=\left( {\Delta x^{\left( t \right)},\Delta y^{\left( t \right)},\Delta h^{\left( t \right)},\Delta \varsigma^{\left( t \right)}} \right)^T\nonumber\\
    &=\left(x-x^{\left( t \right)},y-y^{\left( t \right)}, h-h^{\left( t \right)},\varsigma-\varsigma^{\left( t \right)} \right)^T.\nonumber
\end{align}

The objective function of $\bm{P3}$, i.e., Eq. \eqref{P3_const1}, is the second-order Taylor expansion of the objective of $\bm{P2\_1}$ at point ${{\bf{u}}^{\left( t \right)}}$, and Constraints \eqref{p3_1}, \eqref{p3_2}, \eqref{p3_3}, \eqref{p3_4}, and \eqref{p3_5} are the first-order Taylor expansion of the constraints in $\bm{P2\_1}$ at point ${{\bf{u}}^{\left( t \right)}}$. 
$\bm{P3}$ is a Quadratic Programming (QP) problem with five linear constraints with respect to ${\bf{\Delta }}{\bf{u}}^{\left( t \right)}$, and so the Active-set method \cite{nocedal2006numerical} is used to derive the optimal solution ${\bf{\Delta }}{\bf{u}}^{\left( t \right)}$ and the corresponding Lagrangian multipliers for $\bm{P3}$, denoted as ${\bf{m}}^{\left( t+1\right)}$, which will be used to estimate the Hessian matrix of the Lagrangian function for $\bm{P2\_1}$ in the next iteration. 

 
After deriving the optimal values ${\bf{\Delta }}{\bf{u}}^{(t)}$ by applying Active-set to solve $\bm{P3}$, ${\bf{u}}^{(t+1)}$ is updated, i.e., ${\bf{u}}^{(t+1)}={\bf{u}}^{(t)}+{\bf{\Delta }}{\bf{u}}^{(t)},$
and a new QP subproblem $\bm{P3}$ is formulated during the $\left(t+1\right)^{th}$ iteration based on ${\bf{u}}^{(t+1)}$. Note that the Hessian matrix of the Lagrangian function for $\bm{P3}$, i.e., ${\bf{H}}^{\left( t+1 \right)}$, has to be recalculated in the $\left(t+1\right)^{th}$ iteration to formulate $\bm{P3}$. In order to reduce the complexity, the Broyden–Fletcher–Goldfarb–Shanno (BFGS) algorithm \cite{nocedal2006numerical} is used to update ${\bf{H}}^{\left( t+1 \right)}$, while keeping it positive definite, i.e.,

\begin{equation}
\resizebox{3in}{!}{$\!{{\bf{H}}^{(t+1)}}\!=\!{{\bf{H}}^{(t)}}\!+\! \frac{{{{\bf{q}}^{(t )}}{{\bf{q}}^{(t)}}^{T}}}{{{{\bf{q}}^{(t)}}^{T}\!\!{\bf{\Delta }}{\bf{u}}^{(t)}}} \!-\!\frac{{{{\bf{H}}^{(t)}}\!{\bf{\Delta }}{{\bf{u}}^{\left(t\right)}}\!{\bf{\Delta }}{{\bf{u}}^{\!\left(t\right)}}^{T}\!{{{\bf{H}}^{(t)}}}}}{{({\bf{\Delta }}{\bf{u}}^{\left(t\right)})^{T}{{\bf{H}}^{(t)}}({\bf{\Delta }}{\bf{u}}^{\left(t\right)})}},$}
\label{eq_hessian_update}
\end{equation}
where \resizebox{3in}{!}{${\bm{q}^{\left({t} \right)}}=\nabla _{\bf{u}}\mathcal{L}\left({\bf{u}}^{\left( t+1 \right)},{{\bf{m}}^{\left( {t + 1} \right)}}\right)\!-\!\nabla _{\bf{u}}\mathcal{L}\left({\bf{u}}^{\left( t \right)},{{\bf{m}}^{\left( {t + 1} \right)}}\right).$}

Similarly, the QP problem constructed in the ${\left( {t + 1} \right)^{th}}$ iteration will be solved by applying Active-set to obtain the optimal value of ${\bf{\Delta }}{\bf{u}}^{(t+1)}$. ${\bf{\Delta }}{\bf{u}}^{(t+1)}$ is used to obtain ${\bf{u}}^{(t+2)}$, which will be used to construct a new $\bm{P3}$ for the next iteration. So, the DBS placement algorithm keeps constructing a new $\bm{P3}$ and updating ${\bf{u}}^{(t)}$ by solving $\bm{P3}$ in each iteration until $\Delta \varsigma^{(t)} \le \nu$, where $\nu$ is a predefined threshold.
The DBS placement algorithm is summarized in Algorithm \ref{alg:dlu}.
\subsection{Summary of BROAD}
The position of the DBS is first initialized (e.g., the DBS is placed over the central of a PoI with the height equal to 50 $m$), and the values of ${\bm{\mathcal{Z}}} = \left\{ {z_1,z_2, \cdots z_{\left| \bm{\mathcal{I}} \right|}} \right\}$ are derived by executing user access control. Then, in each iteration, BROAD first executes Algorithm \ref{alg:dlu} to update the position of the DBS based on the values of ${\bm{\mathcal{Z}}}$ generated in the previous iteration and then updates ${\bm{\mathcal{Z}}}$ based on the new position of the DBS by executing user access control. The iteration continuous until the number of the satisfied users (i.e., \resizebox{!}{0.09in}{$\sum\limits_{i \in \bm{\mathcal{I}}} {{z_i}}$}) in the current iteration does not increase as compared to the previous iteration. BROAD is summarized in Algorithm \ref{alg:BROAD}.

\begin{figure}[!t]
 \removelatexerror
\begin{algorithm}[H]
\label{alg:dlu}
\SetAlgoLined
\caption{DBS placement}
Initialize ${\bf{u}}^{(0)}=\left( {{x^{(0)}},{y^{(0)}},{h^{(0)}},{\varsigma^{(0)}}} \right)$, ${\bf{m}}^{(0)}$, ${\bf{\Delta }}{\bf{u}}^{(0)}=inf$, $t=0$.

Calculate ${\bf{H}}^{\left( 0 \right)}$.

\While{True}
{
Construct a quadratic sub-problem $\bm{P3}$;

Derive ${\bf{\Delta }}{\bf{u}}^{(t)}$ and ${{{\bf{m}}^{\left( {t + 1} \right)}}}$ by solving $\bm{P3}$ based on Active-set;

\If{$\Delta \varsigma^{(t)}>\nu$}{break;}


${\bf{u}}^{(t+1)}={\bf{u}}^{(t)}+{\bf{\Delta }}{\bf{u}}^{(t)}$;

Update ${\bf{H}}^{\left( t+1 \right)}$ based on Eq. \eqref{eq_hessian_update};

$t=t+1$;
}
\end{algorithm}
\end{figure}

\begin{figure}[!t]
 \removelatexerror
\begin{algorithm}[H]
\label{alg:BROAD}
\SetAlgoLined
\caption{BROAD}
Initialize the position of the DBS $\left( {x,y,h} \right)$.

Derive ${\bm{\mathcal{Z}}} = \left\{ {z_1,z_2, \cdots z_{\left| \bm{\mathcal{I}} \right|}} \right\}$ based on user access control algorithm.

$g=\sum\limits_{i \in \bm{\mathcal{I}}} {{z_i}}$ and $g^{opt}=0$.

\While{$g>g^{opt}$}
{$g^{opt}=g$;

$x^{opt}=x$; $y^{opt}=y$; $h^{opt}=h$; ${\bm{\mathcal{Z}}}^{opt}={\bm{\mathcal{Z}}}$;

Update the position of the DBS $\left( {x,y,h} \right)$ based on Algorithm \ref{alg:dlu};

Derive ${\bm{\mathcal{Z}}}$ based on user access control algorithm;

$g=\sum\limits_{i \in \bm{\mathcal{I}}} {{z_i}}$;
}
\end{algorithm}
\end{figure}

\section{Simulation}
\begin{figure}[!htb]
	\centering	
        \includegraphics[width=0.8\columnwidth]{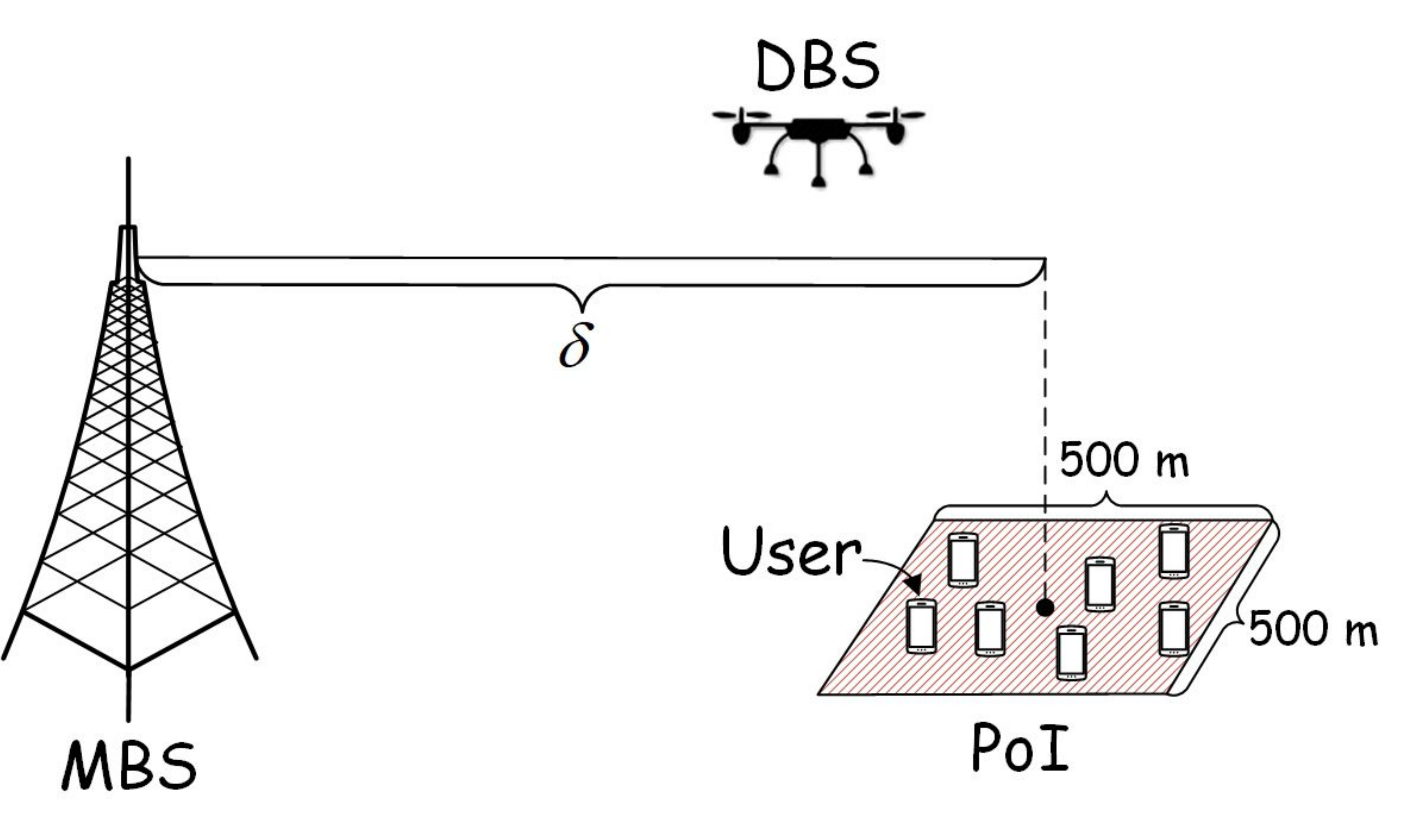} 
	\caption{Simulation setup.}
	\label{fig:ground}
\end{figure}

We conduct extensive simulations to analyze the performance of BROAD. Assume that there are 500 users uniformly distributed in a PoI with the size of 500 m $\times$ 500 m. As shown in Fig. \ref{fig:ground}, the horizontal distance between the center of the PoI and the MBS, denoted as $\delta$, is varied from 5 km to 20 km in the simulation. Data rate requirements of the users (i.e., $\varphi_i$) are generated according to a truncated exponential distribution, where $\varphi_i<500$ Kbps and $\varphi_i>500$ Mbps are truncated. 
Since the users in the PoI are far away from the MBS, they can only communicate with the DBS. The DBS acts as a relay node to forward data from the remote MBS to the users.
Other simulation parameters are listed in Table \ref{tab:sim_para}.

\begin{table}[!htb]
	\centering
	\caption{Simulation Parameters}
	\resizebox{\columnwidth}{!}{%
	\begin{tabular}{ll}
		\toprule
		\textbf{Parameter} & \textbf{Value} \\
		\midrule
		Transmission power of FSO transmitter ($p^{fso}$) & 1 mW \\
		Optical efficiency of FSO transmitter/receiver ($\tau^{tx}$/$\tau^{rx}$) & 0.9/0.7 \\
		Atmospheric attenuation factor ($\gamma$) & 1 dB$/$km \\
		Aperture diameter for the FSO receiver ($\Omega$) & 42.5 mm \\
		Wavelength of the FSO beam ($\iota$) & 1550 nm \\
		Planck constant ($\kappa$) & 6.626$\times$$10^{-34}$ m$^2$kg/s \\
	    Sensitivity of the FSO receiver ($N_b$) & 67885 photons$/$bit \\
		Transmission power of access link ($p^d$)& $0.1$ W\\
		Background noise spectral density($N_0$) & $-104$ dBm/10 MHz\\
		Access link bandwidth ($B$)& 20 MHz\\
        Height of the MBS ($h^{m}$)& 20m\\
        Minimum height of the DBS ($h^{min}$) & 50m\\
        Environmental parameters in Eq. \eqref{eq:los_prob} &$\alpha=9.6, \beta=0.28$\\
		\bottomrule
	\end{tabular}%
	}
	\label{tab:sim_para}%
\end{table}%

To demonstrate the performance of BROAD, the four other existing DBS placement algorithms, i.e., Spectral efficienT Aware DBS pLacement and usEr association (STABLE)\cite{sun2018jointly}, SimultaneOus user Association and DBS Placement (SOAP) \cite{esrafilian2018simultaneous}, SpecTrum efficiency Aware DBS placement and useR association (STAR)\cite{8937522}, and QoS awaRe dronE base Station plaCement and mobile User association stratEgy (RESCUE) \cite{Wu:2019:FDA} are used as the reference algorithms. Here, STABLE, SOAP, and STAR have the same goal, i.e., to optimize the 3D position of the DBS and user association such that the overall SE of the access network is maximized. Still, they apply different methods to calculate the optimal altitude of the DBS. In particular, once the horizontal position of the DBS is calculated, SOAP derives the optimal altitude of the DBS that can maximize the SE of the worst user in the PoI; the optimal altitude of the DBS in STABLE equals the average of the optimal altitudes of the DBS with respect to all the users (for example, if there are two users in the PoI and the optimal altitudes of the DBS with respect to the two users are 100 m and 200 m, respectively, then the optimal altitude of the DBS is 150 m); STAR obtains the optimal altitude of the DBS by formulating an optimization problem (which maximizes the sum of the SE from the DBS to all the users in the PoI) and applying the Projected Gradient Descent method to solve the optimization problem. Note that STABLE, SOAP, and STAR assume that the backhaul link always has enough capacity to satisfy the data rate requirements of the access network. RESCUE aims to jointly optimize the DBS deployment, user association, and bandwidth allocation such that the number of satisfied users is maximized. Also, RESCUE considers the backhaul link capacity to be no less than the sum of the access link data rates for the DBS. The basic idea of RESCUE is to divide the PoI into a number of blocks with the same size, and then apply brute-force search to iteratively evaluate the performance of each block, i.e., what would be the number of satisfied users if the DBS is placed over the center of a block. The optimal 3D position of the DBS would be the block that has the maximum number of satisfied users. The complexity of RESCUE is extremely high due to its brute-force search nature. To reduce the complexity, RESCUE can divide the PoI with larger blocks, which, however, would significantly reduce its performance. In addition, RESCUE assumes the DBS can only be placed over the area within the PoI, which leads to an unsuitable DBS placement when the distance between the PoI and the MBS is large.

\begin{figure}[!htb]
	\centering	
	\includegraphics[width=1.0\linewidth]{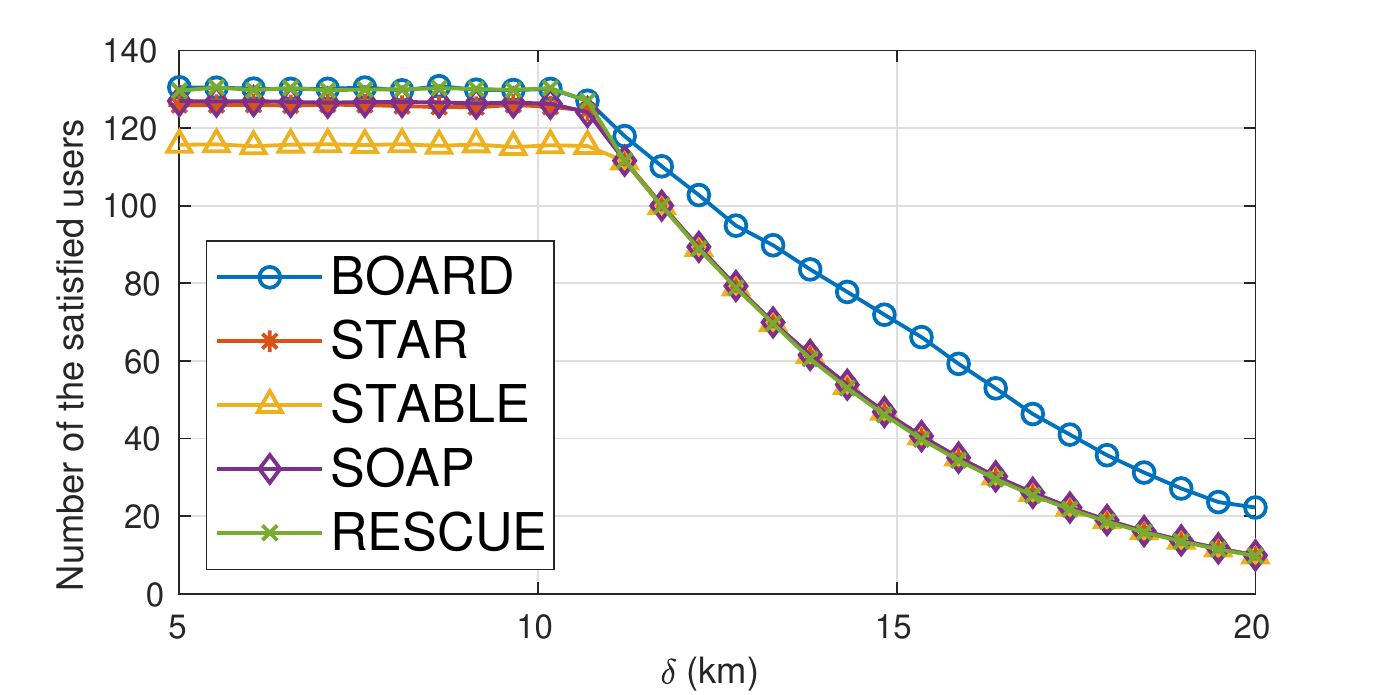}
	\caption{Number of satisfied users over $\delta$.}	
	\label{fig:obj_delta}
\end{figure}

\begin{figure}[!htb]
	\centering	
	\includegraphics[width=1.0\linewidth]{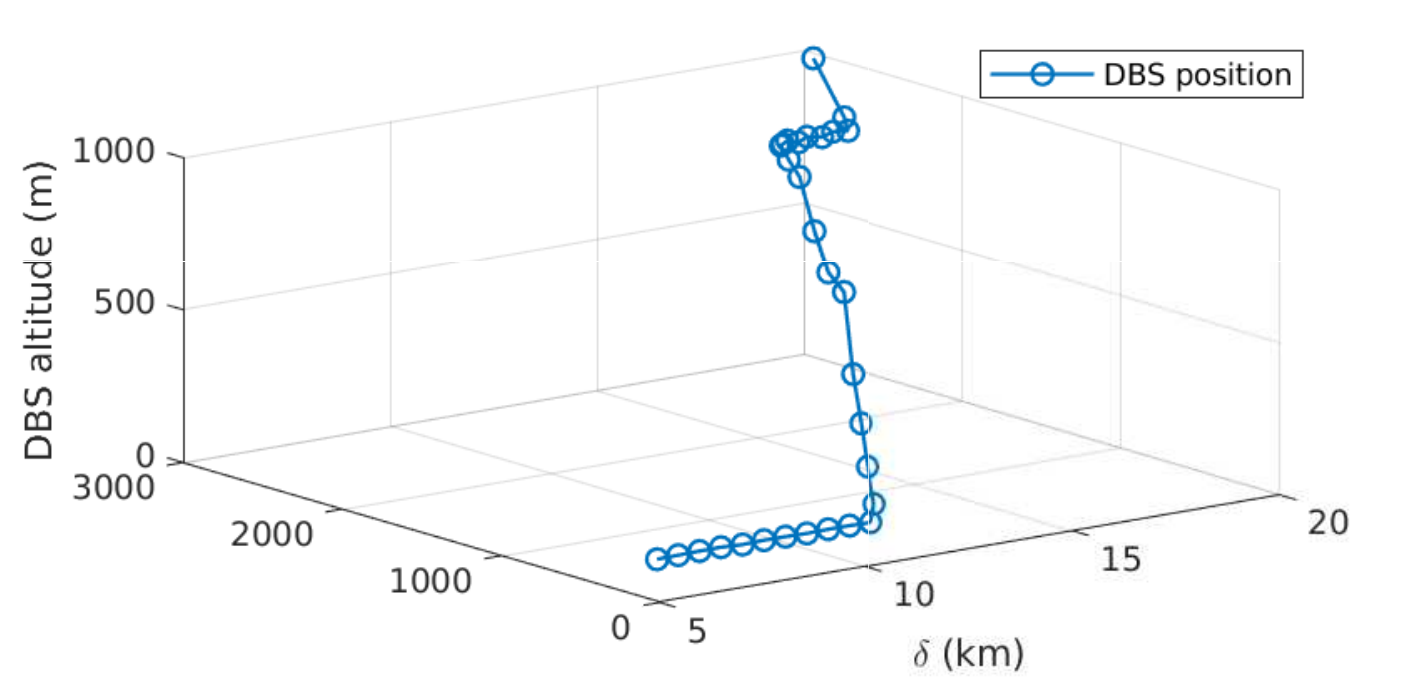}
	\caption{DBS position over $\delta$.}	
	\label{fig:position_delta}
\end{figure}

\begin{figure}[!htb]
	\centering	
	\includegraphics[width=1.0\columnwidth]{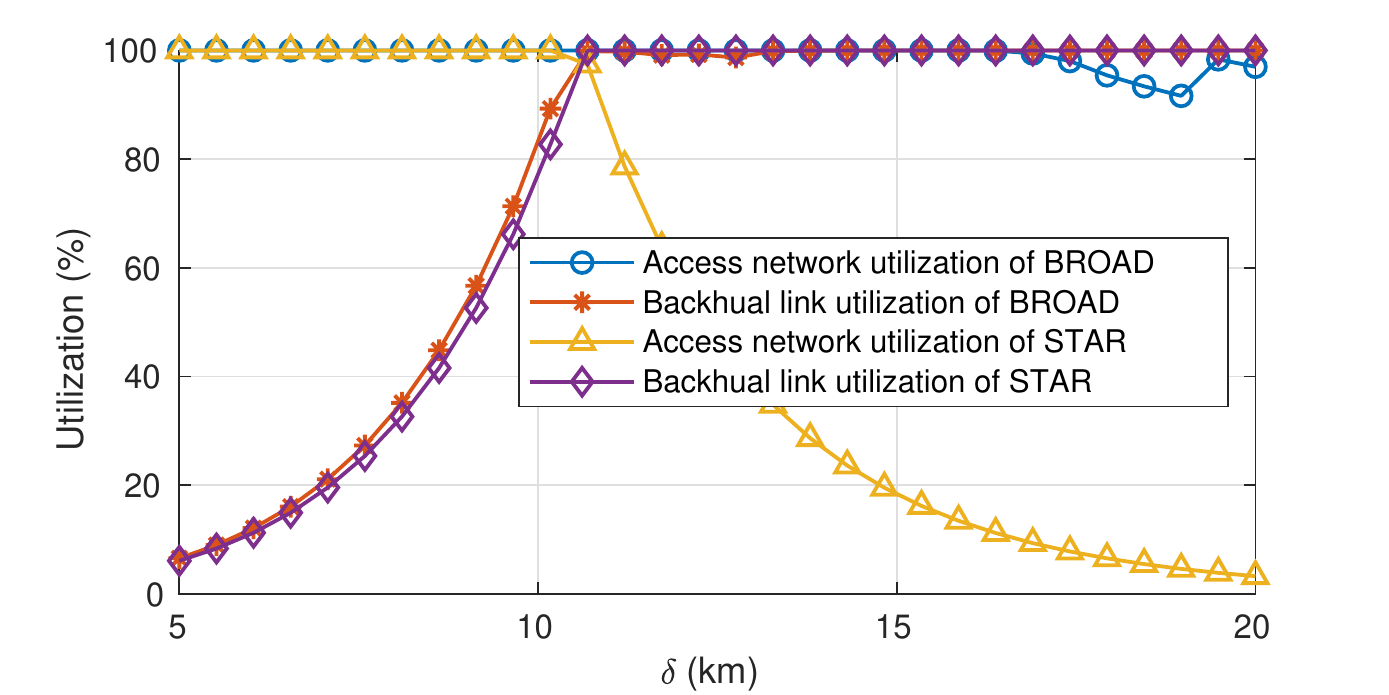}
	\caption{Utilization over $\delta$.}	
	\label{fig:utilization_delta}
\end{figure}

\subsection{Performance analysis over $\delta$}
The horizontal distance $\delta$ between the center of the PoI and the MBS may affect the performance of BROAD and the four reference algorithms. Fig. \ref{fig:obj_delta} shows the number of satisfied users achieved by different algorithms over $\delta$. From the figure, we can see that the performance of BROAD is similar to STAR, SOAP, and RESCUE when $\delta<10$ km. This is because when $\delta<10$ km, the FSO-based backhaul link has sufficient capacity to meet the data rate requirements of the access networks, and so the optimal DBS placement should be over the PoI area. Thus, STAR and STABLE, which optimize the position of the DBS that only maximizes the performance of the access network, have a similar performance to BROAD.

Yet, when $\delta>10$ km, although the number of satisfied users incurred by all the algorithms reduces as $\delta$ increases, BROAD always incurs the most satisfied users because STAR, SOAP, and STABLE do not take the backhaul link capacity constraint into account and place the DBS at inappropriate positions. Specifically, these algorithms deploy their DBSs around the center of the PoI with different altitudes, and the positions of their DBSs do not significantly change as $\delta$ increases, which incurs extreme long distance to the MBS when $\delta>10$ km and leads to the bottleneck on the FSO-based backhaul link, and thus significantly reduce the number of satisfied users. Although RESCUE considers backhaul capacity, it only searches the whole positions within the PoI. Yet, the optimal position of the DBS is out of the PoI when $\delta>10$. On the other hand, BROAD can dynamically adjust the DBS placement by jointly considering the capacities of the FSO-based backhaul link and the access network. Fig. \ref{fig:position_delta} shows the position of the DBS for BROAD over $\delta$, where the x-axis is the value of $\delta$, and the y-axis indicates the horizontal distance between the DBS and the center of the PoI (note that a larger value in the y-axis indicates a shorter distance between the DBS and the MBS), and the z-axis implies the altitude of the DBS. When $\delta>10$ km, BROAD places the DBS further away from the PoI as $\delta$ increases, which implies that when the backhaul link becomes the bottleneck, BROAD prefers to place the DBS to improve the FSO-based backhaul link capacity by sacrificing the performance of the access network. 
Note that, as shown in Fig. \ref{fig:position_delta}, when $\delta>10$ km, the altitude of the DBS generated by BROAD increases as $\delta$ increases. This is because the horizontal distances between the DBS and the users in the PoI increase as the DBS is placed further away from the PoI. So the DBS has to increase its altitude to avoid decreasing the probability of having NLoS to the users and thus avoid the significant decrements of the access link capacities from the DBS to the users.

Fig. \ref{fig:utilization_delta} shows the FSO-based backhaul link utilization (i.e., $\!\sum\limits_{i \in {{\bm{\mathcal{I}}}}}\!\!{\frac{{{z_i}{\varphi _i}}}{{{r^{fso}}}}}$) and the access network bandwidth utilization (i.e., $\sum\limits_{i \in {{\bm{\mathcal{I}}}}}\!\!{\frac{{{z_i}{b_i}}}{B}}\!$) for BROAD and STAR, respectively. Note that the figure does not show the utilization for SOAP, STABLE, and RESCUE since they are similar to the utilization for STAR. From the figure, we can see that, when $\delta>10$ km, STAR incurs a remarkable reduction of the access network bandwidth utilization as $\delta$ increases. This is because STAR does not consider the backhaul link capacity and always deploys the DBS around the center of the PoI, and so the backhaul link capacity significantly reduces as $\delta$ increases. As a result, due to the limited backhaul link capacity,  fewer satisfied users are connected to the DBS, which decreases the amount of bandwidth allocated to the satisfied users, thus significantly reducing the access network bandwidth utilization. On the other hand, the DBS placement incurred by BROAD leads to a more balanced utilization between the FSO-based backhaul link and the access network, i.e., both of them are above $90\%$. A DBS placement with a more balanced utilization results in higher throughput of the path from the MBS to the users via the DBS, thus increasing the number of satisfied users.

\begin{figure}[!htb]
	\centering	
	\includegraphics[width=1.0\columnwidth]{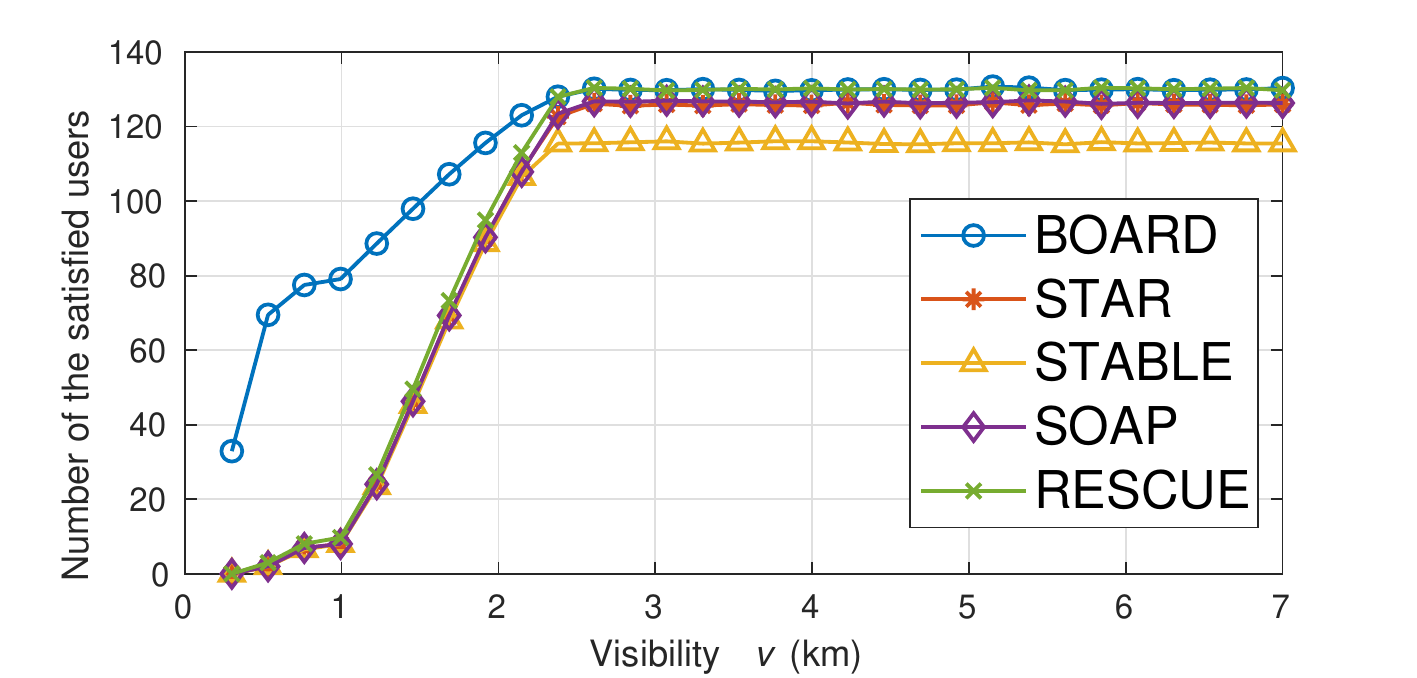}
	\caption{Number of satisfied users over $v$.}
	\label{fig:obj_v}
\end{figure}

\begin{figure}[!htb]
	\centering	
	\includegraphics[width=1.0\linewidth]{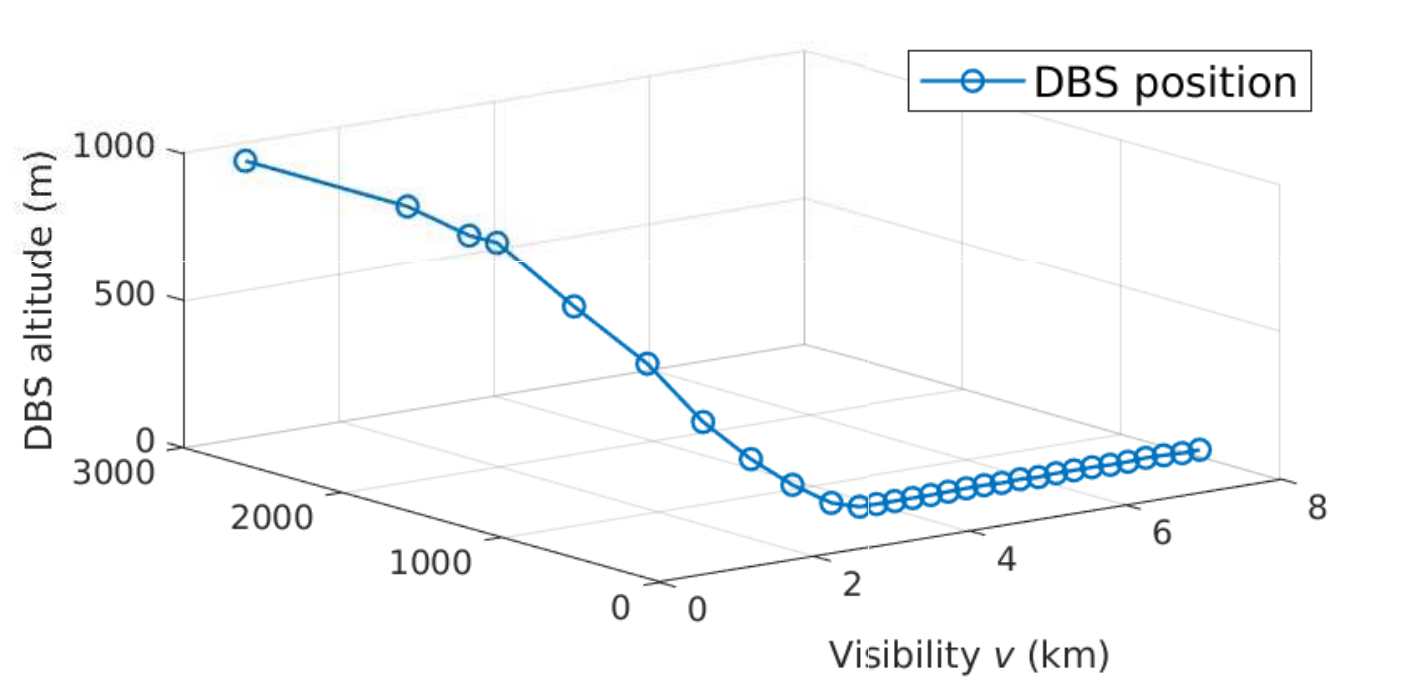}
	\caption{DBS position over $v$.}	
	\label{fig:position_v}
\end{figure}

\begin{figure}[!htb]
	\centering	
	\includegraphics[width=1.0\columnwidth]{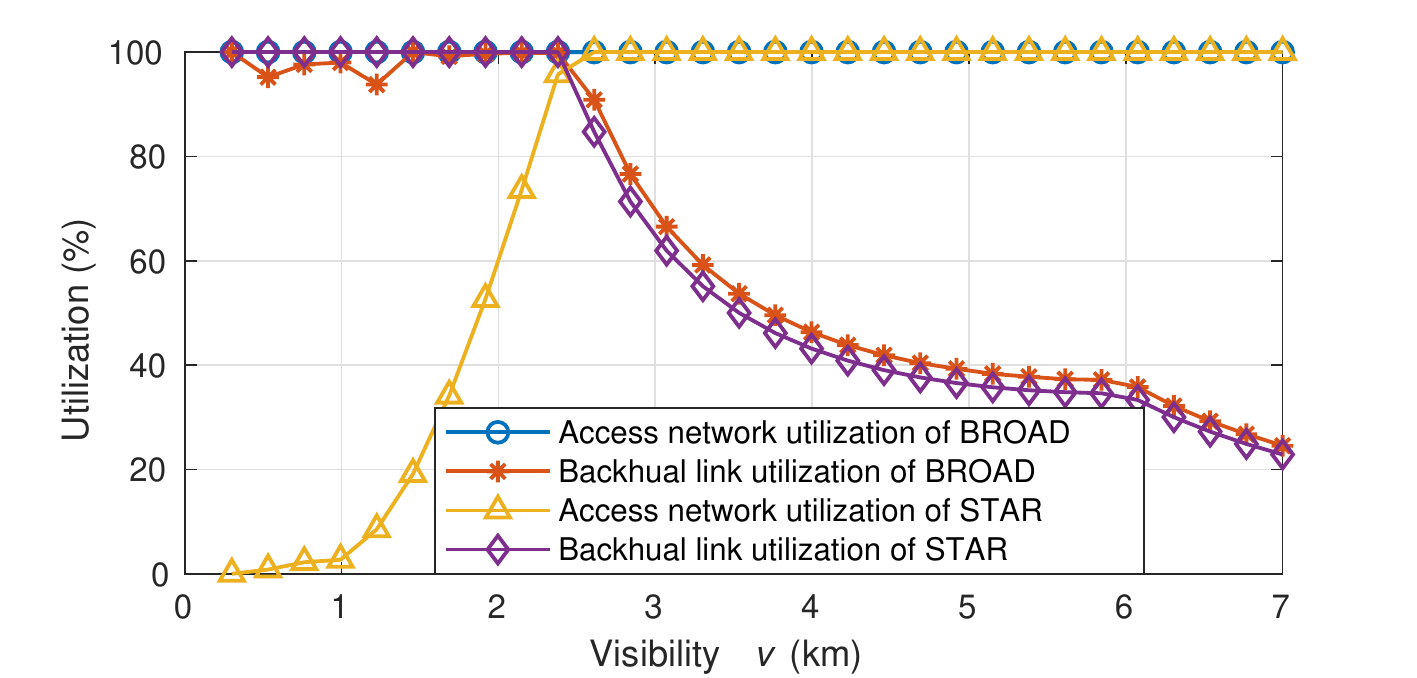}
	\caption{Utilization over $v$.}
	\label{fig:utilization_v}
\end{figure}

\subsection{Performance analysis over visibility distance $v$}
From Eq. \eqref{eq:visibiltity_range}, we can derive that the visibility distance $v$ determines the atmospheric attenuation, which decides the FSO-based backhaul link capacity. In this section, we will analyze how $v$ affects the performance of different algorithms. Note that a larger value of $v$ leads to a higher FSO-based backhaul link capacity. Assume that the distance between the MBS and the center of the PoI is 5 km. Fig. \ref{fig:obj_v} shows the number of satisfied users incurred by different algorithms as $v$ increases. BROAD outperforms other algorithms, especially when the visibility distance $v<2.3$ km. This is because when $v$ reduces, the FSO beam may suffer from more severe attenuation per kilometer, thus reducing the FSO link capacity. As shown in Fig. \ref{fig:position_v}, BROAD can dynamically adjust the DBS placement by reducing the horizontal distance between the MBS and the DBS to maintain the FSO link capacity when $v$ reduces. Yet, the objective of STAR, SOAP, and STABLE is to maximize the performance of the access network without considering the backhaul link capacity. Therefore, as mentioned before, the DBS positions generated by these algorithms are always near the center of the PoI under different values of $v$, which leads to the bottleneck on the FSO-based backhaul link when $v<2.3$ km, thus limiting the number of the satisfied users. From Fig. \ref{fig:utilization_v}, we can see that, in STAR, the access network bandwidth is underutilized, but the FSO-based backhaul link is fully utilized when $v<2.3$ km. On the other hand, BROAD prefers to balance the utilization between the access network and backhaul link when $v<2.3$ km, resulting in more satisfied users. RESCUE, as shown in Fig. \ref{fig:obj_v}, has the similar performance when $v\ge 2.3$ km but worse performance when $v<2.3$ km as compared to BOARD. This is because the optimal position of DBS is out of the PoI when $v<2.3$, but RESCUE can only search the positions within the PoI, thus leading to unappropriated DBS placement.                 

\section{Conclusion}
In this paper, the FSO-based drone-assisted mobile network architecture is proposed, where FSO is applied as the backhauling solution to potentially increase the capacity of the backhaul link. The joint bandwidth allocation and backhaul-aware DBS placement problem are formulated. BOARD algorithm is designed to derive the suboptimal DBS placement and bandwidth allocation such that the number of satisfied users can be maximized. Finally, as compared to the reference algorithms, i.e., STAR, SOAP, and STABLE, the superior performance of BOARD is validated via Monte Carlo simulations.

Achieving FSO-based drone-assisted mobile networks can provide responsive and high-capacity wireless networks for emergency communications. However, there are some challenges in deploying FSO-based drone-assisted mobile networks. First, a DBS could change its location and vibrate freely in the air. Hence, it's difficult to ensure that the optical beam is well-aligned between the FSO transmitter at the MBS and the FSO receiver at the DBS. 
Many high-accuracy ATP systems have been designed to provide high-precise alignment. Yet, these ATP systems are heavy and need to be mounted on both the MBS and the DBS, thus leading to the second challenge,
i.e., short operation time. The current commercial drones are normally powered by portable batteries, which can allow the drones to fly for around 30 min. The flight time would be significantly reduced if the drone is equipped with an FSO transceiver and the corresponding ATP system, which substantially increases the payload and energy consumption of the drone. Some potential solutions to extend the battery life of a DBS include 
1) new battery technologies. Hydrogen fuel cells can offer three times the flight time compared to lithium polymer batteries; 
2) simultaneous Wireless Information and Power Transfer (SWIPT) for FSO \cite{8938182}. The SWIPT technology has been widely adopted in the RF domain to simultaneously transmit both data and energy to a target device. The same idea can be applied to the FSO-based backhaul link, where the MBS can transmit a high-power optical beam, which carries both power and data, to the drone. The drone utilizes part of the received optical beam for data demodulation and the rest of the received optical beam to charge its battery, thus prolonging its battery life.
\bibliographystyle{IEEEtran}
\bibliography{IEEEabrv,mybibliography}

\begin{thebibliography}{10}
\providecommand{\url}[1]{#1}
\csname url@samestyle\endcsname
\providecommand{\newblock}{\relax}
\providecommand{\bibinfo}[2]{#2}
\providecommand{\BIBentrySTDinterwordspacing}{\spaceskip=0pt\relax}
\providecommand{\BIBentryALTinterwordstretchfactor}{4}
\providecommand{\BIBentryALTinterwordspacing}{\spaceskip=\fontdimen2\font plus
\BIBentryALTinterwordstretchfactor\fontdimen3\font minus
  \fontdimen4\font\relax}
\providecommand{\BIBforeignlanguage}[2]{{%
\expandafter\ifx\csname l@#1\endcsname\relax
\typeout{** WARNING: IEEEtran.bst: No hyphenation pattern has been}%
\typeout{** loaded for the language `#1'. Using the pattern for}%
\typeout{** the default language instead.}%
\else
\language=\csname l@#1\endcsname
\fi
#2}}
\providecommand{\BIBdecl}{\relax}
\BIBdecl

\bibitem{Sun:MEC:2017}
N.~Ansari and X.~Sun, ``Mobile edge computing empowers internet of things,''
  \emph{IEICE Trans. Commun.}, vol. 101, no.~3, pp. 604--619, 2018.

\bibitem{Zhang:2018:PMD}
S.~{Zhang}, X.~{Sun}, and N.~{Ansari}, ``Placing multiple drone base stations
  in hotspots,'' in \emph{2018 IEEE 39th Sarnoff Symposium}, Sep. 2018, pp.
  1--6.

\bibitem{Chowdhery:2018:ACP}
A.~{Chowdhery} and K.~{Jamieson}, ``Aerial channel prediction and user
  scheduling in mobile drone hotspots,'' \emph{IEEE/ACM Transactions on
  Networking}, vol.~26, no.~6, pp. 2679--2692, Dec 2018.

\bibitem{Sun:2017:LAD}
X.~{Sun} and N.~{Ansari}, ``Latency aware drone base station placement in
  heterogeneous networks,'' in \emph{2017 IEEE Global Commun. Conf.}, Dec 2017,
  pp. 1--6.

\bibitem{wu2019cooperative}
D.~Wu, X.~Sun, and N.~Ansari, ``A cooperative drone assisted mobile access
  network for disaster emergency communications,'' in \emph{2019 IEEE Global
  Communications Conference (GLOBECOM)}, 2019, pp. 1--6.

\bibitem{8938182}
N.~Ansari, Q.~Fan, X.~Sun, and L.~Zhang, ``Soarnet,'' \emph{IEEE Wireless
  Communications}, vol.~26, no.~6, pp. 37--43, 2019.

\bibitem{Niu:2017:ESB}
Y.~{Niu}, C.~{Gao}, Y.~{Li}, L.~{Su}, D.~{Jin}, Y.~{Zhu}, and D.~O. {Wu},
  ``Energy-efficient scheduling for mmwave backhauling of small cells in
  heterogeneous cellular networks,'' \emph{IEEE Transactions on Vehicular
  Technology}, vol.~66, no.~3, pp. 2674--2687, 2017.

\bibitem{Chehri:2020:PMP}
A.~{Chehri} and H.~T. {Mouftah}, ``Phy-mac mimo precoder design for sub-6 ghz
  backhaul small cell,'' in \emph{2020 IEEE 91st Vehicular Technology
  Conference (VTC2020-Spring)}, 2020, pp. 1--5.

\bibitem{Coldrey:2013:NSC}
M.~{Coldrey}, J.~{Berg}, L.~{Manholm}, C.~{Larsson}, and J.~{Hansryd},
  ``Non-line-of-sight small cell backhauling using microwave technology,''
  \emph{IEEE Communications Magazine}, vol.~51, no.~9, pp. 78--84, 2013.

\bibitem{Siddique:2015:WBS}
U.~{Siddique}, H.~{Tabassum}, E.~{Hossain}, and D.~I. {Kim}, ``Wireless
  backhauling of 5g small cells: challenges and solution approaches,''
  \emph{IEEE Wireless Communications}, vol.~22, no.~5, pp. 22--31, October
  2015.

\bibitem{9540913}
T.~Zhang, X.~Sun, and C.~Wang, ``On optimizing the divergence angle of an
  fso-based fronthaul link in drone-assisted mobile networks,'' \emph{IEEE
  Internet of Things Journal}, vol.~9, no.~9, pp. 6914--6921, 2022.

\bibitem{9302228}
A.~Tosun, H.~Nouri, and M.~Uysal, ``Experimental investigation of pointing
  errors on drone-based fso systems,'' in \emph{28th Signal Processing and
  Communications Applications Conference}, 2020, pp. 1--4.

\bibitem{9557287}
P.~V. Trinh, A.~Carrasco-Casado, T.~Okura, H.~Tsuji, D.~R. Kolev, K.~Shiratama,
  Y.~Munemasa, and M.~Toyoshima, ``Experimental channel statistics of
  drone-to-ground retro-reflected fso links with fine-tracking systems,''
  \emph{IEEE Access}, vol.~9, pp. 137\,148--137\,164, 2021.

\bibitem{Ciaramella:2009:FSO}
E.~{Ciaramella}, Y.~{Arimoto}, G.~{Contestabile}, M.~{Presi}, A.~{D'Errico},
  V.~{Guarino}, and M.~{Matsumoto}, ``1.28-{Tb/s} (32 $\times$ 40 {Gb/s})
  free-space optical wdm transmission system,'' \emph{IEEE Photonics Technology
  Letters}, vol.~21, no.~16, pp. 1121--1123, Aug 2009.

\bibitem{Curran:2017:FSONet}
M.~Curran, M.~S. Rahman, H.~Gupta, K.~Zheng, J.~Longtin, S.~R. Das, and
  T.~Mohamed, ``{FSONet}: A wireless backhaul for multi-gigabit picocells using
  steerable free space optics,'' in \emph{Proceedings of the 23rd Annual
  International Conference on Mobile Computing and Networking}.\hskip 1em plus
  0.5em minus 0.4em\relax New York, NY, USA: ACM, 2017, pp. 154--166.

\bibitem{Esmail:2017:IDH}
M.~A. {Esmail}, A.~{Ragheb}, H.~{Fathallah}, and M.~{Alouini}, ``Investigation
  and demonstration of high speed full-optical hybrid {FSO}/fiber communication
  system under light sand storm condition,'' \emph{IEEE Photonics Journal},
  vol.~9, no.~1, pp. 1--12, Feb 2017.

\bibitem{Zhang:2016:FLU}
H.~{Zhang}, Y.~{Dong}, J.~{Cheng}, M.~J. {Hossain}, and V.~C.~M. {Leung},
  ``Fronthauling for {5G} {LTE-U} ultra dense cloud small cell networks,''
  \emph{IEEE Wireless Commun.}, vol.~23, no.~6, pp. 48--53, December 2016.

\bibitem{Kaushal:2017:OCS}
H.~{Kaushal} and G.~{Kaddoum}, ``Optical communication in space: Challenges and
  mitigation techniques,'' \emph{IEEE Communications Surveys Tutorials},
  vol.~19, no.~1, pp. 57--96, Firstquarter 2017.

\bibitem{AMIRABADI2021165883}
\BIBentryALTinterwordspacing
M.~A. Amirabadi and V.~{Tabataba Vakili}, ``On the performance of a novel
  multi-hop relay-assisted hybrid fso / rf communication system with receive
  diversity,'' \emph{Optik}, vol. 226, p. 165883, 2021. [Online]. Available:
  \url{https://www.sciencedirect.com/science/article/pii/S0030402620317009}
\BIBentrySTDinterwordspacing

\bibitem{Trigui:2017:IMM}
I.~{Trigui}, N.~{Cherif}, S.~{Affes}, X.~{Wang}, V.~{Leung}, and
  A.~{Stephenne}, ``Interference-limited mixed {Málaga-M} and {generalized-K}
  dual-hop {FSO/RF} systems,'' in \emph{2017 IEEE 28th Annual International
  Symposium on Personal, Indoor, and Mobile Radio Communications (PIMRC)}, Oct
  2017, pp. 1--6.

\bibitem{Kaine:1995:IIT}
\BIBentryALTinterwordspacing
M.~Kaine-Krolak and M.~E. Novak, ``An introduction to infrared technology:
  Applications in the home, classroom, workplace, and beyond.'' 1995. [Online].
  Available: \url{https://park.org/Guests/Trace/pavilion/paper11.htm.}
\BIBentrySTDinterwordspacing

\bibitem{Aveta:2018:ICA}
F.~{Aveta}, H.~H. {Refai}, P.~{LoPresti}, S.~A. {Tedder}, and B.~L.
  {Schoenholz}, ``Independent component analysis for processing optical signals
  in support of multi-user communication,'' in \emph{Free-Space Laser
  Communication and Atmospheric Propagation XXX}, vol. 10524.\hskip 1em plus
  0.5em minus 0.4em\relax International Society for Optics and Photonics, 2018,
  p. 105241D.

\bibitem{Zhang:2019:DBS}
S.~Zhang and N.~Ansari, ``3d drone base station placement and resource
  allocation with fso-based backhaul in hotspots,'' \emph{IEEE Transactions on
  Vehicular Technology}, vol.~69, no.~3, pp. 3322--3329, 2020.

\bibitem{Wang:2019:ADU}
Z.~{Wang}, L.~{Duan}, and R.~{Zhang}, ``Adaptive deployment for uav-aided
  communication networks,'' \emph{IEEE Transactions on Wireless
  Communications}, vol.~18, no.~9, pp. 4531--4543, Sep. 2019.

\bibitem{Wu:2018:CTM}
Q.~{Wu} and R.~{Zhang}, ``Common throughput maximization in uav-enabled ofdma
  systems with delay consideration,'' \emph{IEEE Transactions on
  Communications}, vol.~66, no.~12, pp. 6614--6627, Dec 2018.

\bibitem{Al-Hourani:2014:OLA}
A.~Al-Hourani, S.~Kandeepan, and S.~Lardner, ``Optimal {LAP} altitude for
  maximum coverage,'' \emph{IEEE Wireless Commun. Lett.}, vol.~3, no.~6, pp.
  569--572, Dec 2014.

\bibitem{Alzenad:2017:PUA}
M.~{Alzenad}, A.~{El-Keyi}, F.~{Lagum}, and H.~{Yanikomeroglu}, ``3-d placement
  of an unmanned aerial vehicle base station (uav-bs) for energy-efficient
  maximal coverage,'' \emph{IEEE Wireless Communications Letters}, vol.~6,
  no.~4, pp. 434--437, Aug 2017.

\bibitem{Arribas:2019:FCT}
E.~{Arribas}, V.~{Mancuso}, and V.~{Cholvi}, ``Fair cellular throughput
  optimization with the aid of coordinated drones,'' in \emph{IEEE INFOCOM 2019
  - IEEE Conference on Computer Communications Workshops (INFOCOM WKSHPS)},
  April 2019, pp. 295--300.

\bibitem{Sun:2019:JOD}
X.~Sun, N.~Ansari, and R.~Fierro, ``Jointly optimized 3d drone mounted base
  station deployment and user association in drone assisted mobile access
  networks,'' \emph{IEEE Trans. Veh. Technol.}, vol.~69, no.~2, pp. 2195--2203,
  2019.

\bibitem{9537707}
S.~Zhang and N.~Ansari, ``Latency aware 3d placement and user association in
  drone-assisted heterogeneous networks with fso-based backhaul,'' \emph{IEEE
  Transactions on Vehicular Technology}, vol.~70, no.~11, pp. 11\,991--12\,000,
  2021.

\bibitem{Wu:2019:FDA}
D.~Wu, X.~Sun, and N.~Ansari, ``An fso-based drone assisted mobile access
  network for emergency communications,'' \emph{IEEE Transactions on Network
  Science and Engineering}, vol.~7, no.~3, pp. 1597--1606, 2020.

\bibitem{Kalantari:2017:BRD}
E.~{Kalantari}, M.~Z. {Shakir}, H.~{Yanikomeroglu}, and A.~{Yongacoglu},
  ``Backhaul-aware robust 3d drone placement in 5g+ wireless networks,'' in
  \emph{2017 IEEE International Conference on Communications Workshops (ICC
  Workshops)}, May 2017, pp. 109--114.

\bibitem{Sun:2018:JOD}
X.~{Sun} and N.~{Ansari}, ``Jointly optimizing drone-mounted base station
  placement and user association in heterogeneous networks,'' in \emph{2018
  IEEE Intl. Conf. Commun. (ICC)}, May 2018, pp. 1--6.

\bibitem{Zhang:2018:DPI}
L.~{Zhang}, Q.~{Fan}, and N.~{Ansari}, ``3-d drone-base-station placement with
  in-band full-duplex communications,'' \emph{IEEE Communications Letters},
  vol.~22, no.~9, pp. 1902--1905, Sep. 2018.

\bibitem{Zhang:2018:MLP}
Q.~{Zhang}, M.~{Mozaffari}, W.~{Saad}, M.~{Bennis}, and M.~{Debbah},
  ``Machine~learning~for~predictive~on--demand~deployment~of~uavs~for~wireless~communications,''
  in \emph{2018 IEEE Global Commun. Conf.}, Dec 2018, pp. 1--6.

\bibitem{Ghanavi:2018:EAB}
R.~{Ghanavi}, E.~{Kalantari}, M.~{Sabbaghian}, H.~{Yanikomeroglu}, and
  A.~{Yongacoglu}, ``Efficient 3d aerial base station placement considering
  users mobility by reinforcement learning,'' in \emph{2018 IEEE Wireless
  Commun. and Netw. Conf. (WCNC)}, April 2018, pp. 1--6.

\bibitem{Khawaja:2019:SAG}
W.~{Khawaja}, I.~{Guvenc}, D.~W. {Matolak}, U.~{Fiebig}, and
  N.~{Schneckenberger}, ``A survey of air-to-ground propagation channel
  modeling for unmanned aerial vehicles,'' \emph{IEEE Commun. Surveys Tuts.},
  pp. 1--1, 2019.

\bibitem{8688470}
X.~Guo, N.~R. Elikplim, N.~Ansari, L.~Li, and L.~Wang, ``Robust wifi
  localization by fusing derivative fingerprints of rss and multiple
  classifiers,'' \emph{IEEE Transactions on Industrial Informatics}, vol.~16,
  no.~5, pp. 3177--3186, 2020.

\bibitem{al2014optimal}
A.~Al-Hourani, S.~Kandeepan, and S.~Lardner, ``Optimal lap altitude for maximum
  coverage,'' \emph{IEEE Wireless Communications Letters}, vol.~3, no.~6, pp.
  569--572, 2014.

\bibitem{Hu:2007:FOC}
G.-y. Hu, C.-y. Chen, and Z.-q. Chen, ``Free-space optical communication using
  visible light,'' \emph{Journal of Zhejiang University-SCIENCE A}, vol.~8,
  no.~2, pp. 186--191, 2007.

\bibitem{6843013}
S.~Singh and G.~Soni, ``Pointing error evaluation in fso link,'' in \emph{Fifth
  International Conference on Advances in Recent Technologies in Communication
  and Computing (ARTCom 2013)}, 2013, pp. 365--370.

\bibitem{Sun:2022:LCA}
X.~Sun, T.~Zhang, S.~Shao, B.~Tice, P.~Tice, and S.~Jayaweera, ``Low cost atp
  system design for free space optics based drone assisted wireless networks,''
  in \emph{2022 IEEE Globecom Workshops}, 2022, in press.

\bibitem{Liu:2020:DED}
H.-Y. Liu, X.-H. Tian, C.~Gu, P.~Fan, X.~Ni, R.~Yang, J.-N. Zhang, M.~Hu,
  J.~Guo, X.~Cao, X.~Hu, G.~Zhao, Y.-Q. Lu, Y.-X. Gong, Z.~Xie, and S.-N. Zhu,
  ``{Drone-based entanglement distribution towards mobile quantum networks},''
  \emph{Natl. Sci. Rev.}, vol.~7, no.~5, pp. 921--928, 2020.

\bibitem{Fernandes:2022:FST}
M.~A. Fernandes, P.~P. Monteiro, and F.~P. Guiomar, ``Free-space terabit
  optical interconnects,'' \emph{Journal of Lightwave Technology}, vol.~40,
  no.~5, pp. 1519--1526, 2022.

\bibitem{4267802}
A.~A. Farid and S.~Hranilovic, ``Outage capacity optimization for free-space
  optical links with pointing errors,'' \emph{Journal of Lightwave Technology},
  vol.~25, no.~7, pp. 1702--1710, 2007.

\bibitem{shah2019genetic}
S.~Shah, ``Genetic algorithm for the 0/1 multidimensional knapsack problem,''
  \emph{arXiv preprint arXiv:1908.08022}, 2019.

\bibitem{chu1998genetic}
P.~C. Chu and J.~E. Beasley, ``A genetic algorithm for the multidimensional
  knapsack problem,'' \emph{Journal of heuristics}, vol.~4, no.~1, pp. 63--86,
  1998.

\bibitem{pirkul1987heuristic}
H.~Pirkul, ``A heuristic solution procedure for the multiconstraint zero-one
  knapsack problem,'' \emph{Naval Research Logistics (NRL)}, vol.~34, no.~2,
  pp. 161--172, 1987.

\bibitem{nocedal2006numerical}
J.~Nocedal and S.~Wright, \emph{Numerical optimization}.\hskip 1em plus 0.5em
  minus 0.4em\relax Springer Science \& Business Media, 2006.

\bibitem{sun2018jointly}
X.~Sun and N.~Ansari, ``Jointly optimizing drone-mounted base station placement
  and user association in heterogeneous networks,'' in \emph{2018 IEEE
  International Conference on Communications (ICC)}.\hskip 1em plus 0.5em minus
  0.4em\relax IEEE, 2018, pp. 1--6.

\bibitem{esrafilian2018simultaneous}
O.~Esrafilian and D.~Gesbert, ``Simultaneous user association and placement in
  multi-uav enabled wireless networks,'' in \emph{WSA 2018; 22nd International
  ITG Workshop on Smart Antennas}.\hskip 1em plus 0.5em minus 0.4em\relax VDE,
  2018, pp. 1--5.

\bibitem{8937522}
X.~Sun, N.~Ansari, and R.~Fierro, ``Jointly optimized 3d drone mounted base
  station deployment and user association in drone assisted mobile access
  networks,'' \emph{IEEE Transactions on Vehicular Technology}, vol.~69, no.~2,
  pp. 2195--2203, 2020.

\end{thebibliography}

\end{document}